\documentclass[prd,floatfix,onecolumn,amsmath,amssymb,floatfix]{revtex4}
\usepackage{graphicx,color,dcolumn,booktabs,bm}
\usepackage{subfigure}
\usepackage{longtable,lscape}
\usepackage{amssymb}
\usepackage{indentfirst}
\usepackage{epsfig}
\usepackage{feynmf}   
\usepackage{epstopdf}   
\usepackage{slashed}  
\usepackage{cases}
\usepackage[pdfpages]{xcolor}
\definecolor{maroon}{RGB}{139,25,150}
\usepackage{multirow}
\usepackage{float}
\usepackage{graphicx,color,dcolumn,booktabs,bm}
\usepackage[colorlinks, citecolor=blue,anchorcolor=red,menucolor=red, linkcolor=red,filecolor=red,runcolor=red,urlcolor=blue,frenchlinks=red, urlcolor=blue]{hyperref}
\usepackage{orcidlink}

\begin{document}

\preprint{}
\preprint{}
\title{\color{maroon}{Semileptonic $\Omega_{b}\rightarrow
\Omega_{c}{\ell}\bar\nu_{\ell}$ transition in full QCD }}

\author{Z. Neishabouri$^{a}$\,\orcidlink{0009-0009-0892-384X}, K. Azizi$^{a,b}$\,\orcidlink{0000-0003-3741-2167}\thanks{Corresponding Author}, H. R. Moshfegh$^{a,c}$\,\orcidlink{0000-0002-9657-7116}}

\affiliation{
		$^{a}$Department of Physics, University of Tehran, North Karegar Avenue, Tehran 14395-547, Iran\\
		$^{b}$Department of Physics, Do\v{g}u\c{s} University, Dudullu-\"{U}mraniye, 34775 Istanbul, T\"urkiye\\
                $^{c}$Departamento de F\'{i}cia Universidade Cat\'{o}lica do Rio de Janerio, Rio de Janerio22452-970, Brazil}

\date{\today}

\begin{abstract}
We investigate the semileptonic decay of $\Omega_b\to\Omega_c~{\ell}\bar\nu_{\ell}$ in three lepton channels. To this end, we use the QCD sum rule method in three point framework to calculate the form factors defining the matrix elements of these transitions. Having calculated the form factors as building blocks, we calculate the decay widths and branching fractions of the exclusive decays in all lepton channels and compare the results with other theoretical predictions. The obtained results for branching ratios and ratio of branching fractions at different leptonic channels may help experimental groups in their search for these weak decays. Comparison of the obtained results with possible future experimental data can be useful to check the order of consistency between the standard model theory predictions and data on the heavy baryon decays.

\end{abstract}

\maketitle
\section{Introduction}
In recent years, there has been considerable attention given to the study of hadrons containing heavy quarks. Following the first observation of deviations from the Standard Model (SM) predictions in $B$ meson decays at different experiments, attention has shifted towards all members of the b-hadrons. The BaBar experiment reported a $3.4\sigma$ deviation from the SM prediction in the semileptonic $B\to D$ decay, specifically in the ratio of branching fractions between the $\tau$ channel and either $e$ or $\mu$ that couldn't be explained \cite{BaBar:2012obs}. Subsequently, the LHCb experiment reported a violation of Lepton Flavor Universality (LFU) in the ratio of branching fractions for semileptonic $B\to K$ decay, observed at the $\mu$ channel compared to $e$, by $2.6\sigma$ from the SM prediction \cite{LHCb:2014vgu}. Furthermore, discrepancies from the SM predictions have been observed in the $B\to J/\Psi$ decay, showing  $2\sigma$ deviations above the range of central values predicted by the SM in the ratio of branching fractions between the $\tau$ and $\mu$ channels \cite{LHCb:2017vlu}.
The semileptonic decays of b-hadrons hold promise for exploring Beyond the Standard Model (BSM) physics. The decay of $\Lambda_b$ to $\Lambda_c$ has been studied using various theoretical methods \cite{Pervin:2005ve, Faustov:2016pal, Gutsche:2014zna, Detmold:2015aaa, Miao:2022bga, Gutsche:2015mxa,Duan:2022uzm, Azizi:2018axf}, but LHCb has not observed any deviations from the SM predictions \cite{LHCb:2022piu}. According to the quark model, which serves as a framework to describe particles in the SM, nine single-heavy spin one-half ground state baryons can be made of a $b$ quark.
All of these baryons together with the spin three-half single heavy baryon members have been detected in the experiment so far. The heaviest particle among the spin one-half single heavy baryons is the $\Omega_b$, containing a bottom and two strange quarks. The first observation of $\Omega_b$ occurred in $p\bar p$ collisions at $\sqrt s = 1.96~TeV$, through the reconstructed decay $\Omega_b\to J/\Psi \Omega^-$ in the D0 detector at Fermilab's Tevatron, with  $J^P=\frac{1}{2}^+$. The signal event at a mass of $6.165 \pm0.010\pm0.013~GeV$ was with a significance of  $5.4\sigma$ \cite{D0:2008sbw}. After measuring mass and lifetime by the Collider Detector at Fermilab \cite{CDF:2009sbo}, its mass and lifetime have then been respectively measured as $6045.1\pm3.2\pm0.5\pm0.6~MeV/c^2$ and $1.78 \pm0.26\pm0.05\pm0.06~ps$  by LHCb \cite{LHCb:2016coe}.

The semileptonic decays of b-baryons can also be served as  good probes to check the SM predictions with the results of ongoing progressive experiments. It is important to check whether there are similar deviations between the SM predictions and experimental data in b-baryon semileptonic decays or not. Such possible deviations will increase our hope to indirectly search for new physics BSM. In this context, we investigate the semileptonic decay of $\Omega_b \to \Omega_c$   using the QCD sum rule (QCDSR) method, which is a powerful tool to study the  non-perturbative phenomena.
The study of weak decays offers two advantages: Firstly, it provides valuable insights into various SM parameters including  the Cabibbo-Kobayashi-Maskawa (CKM) matrix elements; secondly, it provides us with insight into the lepton flavor universality and serves as a good candidate for studying BSM physics. 
Since the SM predicts the same coupling to the $W$ and $Z$ gauge bosons for all three lepton families, measuring the ratio of branching fraction:
\begin{equation}\label{MuonElectronRatio}
R_{\Omega_c}=\frac{Br[\Omega_b^-\rightarrow \Omega_c\tau\bar\nu_\tau]}{Br[\Omega_b^-\rightarrow \Omega_c (e,\mu)\bar\nu_{(e,\mu)}]},
\end{equation}
and comparing it with future experimental data can open a new window for exploring LFU.

The weak semileptonic $\Omega_b \to \Omega_c$   channel has not been observed yet, but some theoretical computations to calculate the corresponding decay rates have been conducted. The semileptonic decay of $\Omega_b \to \Omega_c$ was investigated by heavy quark effective theory (HQET)\cite{Xu:1992hj}, 1/m corrections to  form factors in the nonrelativistic quark model \cite{Cheng:1995fe},
 the spectator quark model \cite{Singleton:1990ye},  relativistic three quark model \cite{Ivanov:1996fj, Ebert:2006hm, Ebert:2008oxa,Ebert:2006rp,Ivanov:1999pz},  constituent quark model \cite{Pervin:2006ie}, an independent method \cite{Sheng:2020drc}, using the Beth-Salpeter equation \cite {Ivanov:1998ya,Rusetsky:1997id},  large $N_c $  method in HQET framework \cite{ Han:2020sag,Du:2011nj},  light front approach \cite{Zhao:2018zcb,Ke:2012wa} and Bjorken sum rules \cite{Xu:1993mj}.
Our aim is to calculate the decay rates and branching ratios of $\Omega_{b}\rightarrow
\Omega_{c}{\ell}\bar\nu_{\ell}$ using the QCDSR method in full theory for the first time. As stated, the QCDSR method is one of the powerful and predictive models in the non-perturbative area of QCD developed by Shifman, Vainshtein and Zakharov \cite{Shifman:1978by,Shifman:1978bx} and it is used to calculate different hadronic parameters.
This method has had good predictions and comparable results with the experiments so far and it is a trustworthy  non-perturbative approach to study the hadronic decays \cite{Shifman:2001ck,Aliev:2010uy,Aliev:2009jt,Aliev:2012ru,Agaev:2016dev,Azizi:2016dhy}.

The manuscript is organized as follows. In section \ref{Sec2}, we obtain the sum rule for the form factors entering the low energy amplitude of the decay under study.
 In section \ref{Sec3}, we conduct a numerical analysis of the form factors by determining the  working regions of   auxiliary parameters and find the fit functions for the behavior of form factors in terms of transferred momentum squared. We determine the decay rates and branching ratios for all the lepton channels, and compare our results with the predictions of other theoretical studies in section \ref{Sec4}. Section \ref{Sec5} is devoted to the concluding remarks. Some details of the calculations are presented in the Appendices.

\section{QCD sum rule Calculations}~\label{Sec2}
QCD sum rule provides two perspectives on a hadron: The physical or phenomenological side, which views the hadron as a unique object in time-like region, and the QCD or theoretical side, which perceives the hadron's content through constituent quarks and gluons and their interactions in space-like region. By connecting these two perspectives through dispersion integrals and the quark-hadron duality assumption, hadronic parameters are derived in terms of fundamental QCD parameters \cite{Shifman:1978by,Shifman:1978bx,Gross:2022hyw,Shifman:2010zzb}. The form factors relevant to the semileptonic decays are obtained by equating the coefficients of the corresponding Lorentz structures from these two parts  and  applying double Borel transformation and continuum subtraction.

\subsection{Phenomenological side}
To obtain the amplitude for the $\Omega_b \to \Omega_c{\ell}\bar\nu_{\ell}$ transition we need the responsible Hamiltonian at quark level. In this decay $s$ quarks are spectators, so transition happens via $b\to c{\ell}\bar\nu_{\ell}$ with transition current of $J^{tr}=\bar c ~\gamma_\mu(1-\gamma_5)~ b$. The effective Hamiltonian can be written as: 

\begin{eqnarray}\label{Heff}
{\cal H}_{eff} =
\frac{G_F}{\sqrt2} V_{cb} ~\bar c \gamma_\mu(1-\gamma_5) b ~\bar{\ell}\gamma^\mu(1-\gamma_5) \nu_{\ell},
\end{eqnarray}
where $G_F$ is the Fermi coupling constant and $V_{cb}$ is the CKM matrix elements. Obtaining the decay amplitude involves sandwiching the effective Hamiltonian between the initial and final baryon states. In this process, the leptonic part exits the matrix element and we have:

\begin{eqnarray}\label{amp}
&&M=\langle \Omega_{c}\vert{\cal H}_{eff}\vert \Omega_{b}\rangle \notag \\
&&=\frac{G_F}{\sqrt2} V_{cb}\bar{\ell}~\gamma^\mu(1-\gamma_5) \nu_{\ell} \langle \Omega_{c}\vert \bar c \gamma_\mu(1-\gamma_5)b\vert \Omega_{b}\rangle.
\end{eqnarray}
The decay contains two parts: the vector ($V^\mu$) and the axial vector ($A^\mu$) transitions.  Each of these parts can be parameterized in terms of three form factors in full QCD. The most complete parameterizations considering the Lorentz invariance and parity are \cite{Azizi:2018axf}: 

\begin{eqnarray}\label{Cur.with FormFac.}
&&\langle \Omega_c(p',s')|V^{\mu}|\Omega_b (p,s)\rangle = \bar
u_{\Omega_c}(p',s') \Big[F_1(q^2)\gamma^{\mu}+F_2(q^2)\frac{p^{\mu}}{m_{\Omega_b}}
+F_3(q^2)\frac{p'^{\mu}}{m_{\Omega_c}}\Big] u_{\Omega_b}(p,s), \notag \\
&&\langle \Omega_c(p',s')|A^{\mu}|\Omega_b (p,s)\rangle = \bar u_{\Omega_c}(p',s') \Big[G_1(q^2)\gamma^{\mu}+G_2(q^2)\frac{p^{\mu}}{m_{\Omega_b}}+G_3(q^2)\frac{p'^{\mu}}{m_{\Omega_c}}\Big]
\gamma_5 u_{\Omega_b}(p,s), \notag \\
\end{eqnarray}
where, $ F_1(q^2),~F_2(q^2),~ F_3(q^2)$  and  $ G_1(q^2),~G_2(q^2),~ G_3(q^2)$ are form factors describing the vector and axial  transitions, respectively. $q=p-p'$ is the momentum transferred to the leptons, and $u_{\Omega_{b}}(p,s)$ and
$u_{\Omega_{c}}(p',s')$ are Dirac spinors of the initial and final baryonic states. To find the form factors, we use
an appreciate three point correlation function. In this framework, the initial hadron can emerge from the vacuum state and subsequently, after interacting with the weak current, the final hadron can be annihilated into the vacuum state:
\begin{eqnarray}\label{CorFunc}
\Pi_{\mu}(p,p^{\prime},q)&=&i^2\int d^{4}x e^{-ip\cdot x}\int d^{4}y e^{ip'\cdot y}  \langle 0|{\cal T}\{{\cal J}^{\Omega_c}(y){\cal
J}_{\mu}^{tr,V(A)}(0) {\cal J}^{\dag \Omega_b}(x)\}|0\rangle,
\end{eqnarray}
where $\cal T$ is the time-ordering operator, and $ {\cal J}^{\Omega_{b}}(x)$ and ${\cal J}^{\Omega_{c}}(y) $ are the initial and final hadron's interpolating currents, respectively. 
To evaluate the correlation function on the phenomenological side, one needs to insert two relevant hadronic complete sets with the same quantum numbers as the currents ${\cal J}^{\Omega_{b}}$ and ${\cal J}^{\Omega_{c}}$ for each the initial and final  hadrons, respectively, as follows:  
\begin{eqnarray} \label{compelet set}
1=\vert 0\rangle\langle0\vert +\sum_{h}\int\frac{d^4p}{(2\pi)^4}(2\pi) \delta(p^2_h-m^2_h)|h(p)\rangle  \langle h(p)|+\mbox{higher Fock states}.
\end{eqnarray}
After performing some algebraic manipulations, the hadronic side for the correlation function is obtained in the following form:
\begin{eqnarray} \label{PhysSide}
\Pi_{\mu}^{Phys.}(p,p',q)=\frac{\langle 0 \mid {\cal J}^{\Omega_c} (0)\mid \Omega_c(p') \rangle \langle \Omega_{c} (p')\mid
{\cal J}_{\mu}^{tr,V(A)}(0)\mid \Omega_b(p) \rangle \langle \Omega_{b}(p)
\mid {\cal J}^{\dag \Omega_b}(0)\mid
0\rangle}{(p'^2-m_{\Omega_c}^2)(p^2-m_{\Omega_b}^2)}+\cdots~,
\end{eqnarray}
where $\cdots$ means the contributions of the higher states and
 continuum. 
 The residues of the initial ($\lambda_{\Omega_b}$) and final ($\lambda_{\Omega_c}$) states are defined as:
\begin{eqnarray}\label{MatrixElements}
&&\langle 0|{\cal J}^{\Omega_c}(0)|\Omega_c(p')\rangle =
\lambda_{\Omega_c} u_{\Omega_c}(p',s'), \notag \\
&&\langle\Omega_b(p)|\bar {\cal J}^{\Omega_b}(0)| 0 \rangle =
\lambda^{+}_{\Omega_b}\bar u_{\Omega_b}(p,s).
\end{eqnarray}
 Now, by using the summations over Dirac spinors:
\begin{eqnarray}\label{Spinors}
\sum_{s'} u_{\Omega_c} (p',s')~\bar{u}_{\Omega_c}
(p',s')&=&\slashed{p}~'+m_{\Omega_c},\notag \\
\sum_{s} u_{\Omega_b}(p,s)~\bar{u}_{\Omega_b}(p,s)&=&\slashed
p+m_{\Omega_b},
\end{eqnarray}
as well as inserting all the matrix elements, defined above,  into Eq.(\ref{PhysSide}),  we can obtain  the final form of the phenomenological side after performing the double Borel transformation:
\begin{eqnarray}\label{Physical Side structures}
&&\mathbf{\widehat{B}}~\Pi_{\mu}^{\mathrm{Phys.}}(p,p',q)=\lambda_{\Omega_b}\lambda_{\Omega_c}~e^{-\frac{m_{\Omega_b}^2}{M^2}}
~e^{-\frac{m_{\Omega_c}^2}{M'^{2}}}\Bigg[F_{1}\bigg(m_{\Omega_b} m_{\Omega_c} \gamma_{\mu}+m_{\Omega_b} \slashed{p}' \gamma_{\mu}+m_{\Omega_c}\gamma_{\mu}\slashed {p}+\slashed {p}'\gamma_\mu\slashed {p}\bigg)+\notag\\
&&F_2\bigg(\frac{m_{\Omega_c}}{m_{\Omega_b}}p_\mu\slashed {p}+\frac{1}{m_{\Omega_b}}p_{\mu}\slashed {p}' \slashed {p}+m_{\Omega_c}p_\mu +p_\mu\slashed {p}'\bigg)+ F_3\bigg(\frac{1}{m_{\Omega_c}} p'_{\mu} \slashed {p}' \slashed{p}+p'_\mu\slashed {p}'+p'_\mu\slashed {p}+m_{\Omega_b}p'_\mu\bigg)-\notag\\
&& G_1\bigg(m_{\Omega_b} m_{\Omega_c} \gamma_{\mu}\gamma_{5}+m_{\Omega_b}\slashed {p}'\gamma_\mu\gamma_5-m_{\Omega_c}\gamma_\mu\slashed {p}\gamma_5-\slashed {p}'\gamma_\mu\slashed {p}\gamma_5\bigg)- G_2\bigg(p_\mu\slashed {p}'\gamma_5+m_{\Omega_c}p_\mu\gamma_5-\frac{m_{\Omega_c}}{m_{\Omega_b}}p_\mu\slashed {p}\gamma_5-\frac{1}{m_{\Omega_b}} p_{\mu} \slashed {p}' \slashed
{p}\gamma_{5}\bigg)-\notag\\
&&G_3\bigg(\frac{m_{\Omega_b}}{m_{\Omega_c}}p'_\mu\slashed {p}'\gamma_5+m_{\Omega_b}p'_\mu\gamma_5-\frac{1}{m_{\Omega_c}} p'_{\mu}
\slashed {p}'\slashed{p}\gamma_{5}-p'_\mu\slashed {p}\gamma_5\bigg)\Bigg]+..., \notag \\
\end{eqnarray}
where $ M^2 $ and $M'^{2}$ are Borel parameters that should be fixed in the numerical analysis section. 
\\
\subsection{QCD side}
To evaluate the correlation function on the QCD side in deep Euclidean region, one should insert the interpolating currents of hadrons into the correlation function, i.e. Eq.\ (\ref{CorFunc}).
The interpolating current of single heavy $\Omega_Q$ baryon with spin-parity $J^P=(\frac{1}{2})^+$ is given by (see also \cite{Agaev:2017jyt}):
%
\begin{eqnarray} \label{Current}
{\cal J}^{\Omega_{Q}}(x)&&=\frac{-1}{2}~\epsilon_{abc}\Bigg\{\Big(q^{aT}(x)CQ^{b}(x)\Big)\gamma_{5}q^{c}(x) + 
\beta\Big(q^{aT}(x)C\gamma_{5}Q^{b}(x)\Big)q^{c}(x)  -\Bigg[\Big(Q^{aT}(x)Cq^{b}(x)\Big)\gamma_{5}q^{c}(x)    \notag \\
&&+\beta\Big(Q^{aT}(x)C\gamma_{5}q^{b}(x)\Big)q^{c}(x)\Bigg]\Bigg\},~
\end{eqnarray}
where $a,~b$ and $c$ are color indices, $C$ is the charge
conjugation operator, $q$ is $s$ quark and $Q$ is bottom or charm 
quark field. The $\beta$ is a general mixing parameter with $\beta=-1$
being corresponding to  Ioffe current. We will also back to fix the working region of this parameter in numerical analysis section. A proof of the above current considering all the quantum numbers and properties of $\Omega_Q$  is given in  Appendix A.
Now, we are in a position to evaluate the correlation function in the QCD side at coordinate space. By replacing the interpolating currents of the initial and final hadrons $({\cal J}^{\Omega_{b}}, {\cal J}^{\Omega_{c}})$, as well as the transition current $(J^{tr})$ of the decay, in the correlation function Eq (\ref{CorFunc}), and considering all possible contractions of the quark fields using Wick's theorem, we find the correlator in terms of the heavy and light quarks' propagators. To avoid the inclusion of a lengthy formula inside the text, we present it in Appendix B.
For the light quark propagator we use \cite{Agaev:2020zad}:
%

\begin{eqnarray}\label{LightProp}
S_{q}^{ab}(x)&=&i\delta _{ab}\frac{\slashed x}{2\pi ^{2}x^{4}}-\delta _{ab}%
\frac{m_{q}}{4\pi ^{2}x^{2}}-\delta _{ab}\frac{\langle\overline{q}q\rangle}{12} +i\delta _{ab}\frac{\slashed xm_{q}\langle \overline{q}q\rangle }{48}%
-\delta _{ab}\frac{x^{2}}{192}\langle \overline{q}g_{}\sigma
Gq\rangle+
i\delta _{ab}\frac{x^{2}\slashed xm_{q}}{1152}\langle \overline{q}g_{}\sigma Gq\rangle \notag\\
&-&i\frac{g_{}G_{ab}^{\alpha \beta }}{32\pi ^{2}x^{2}}\left[ \slashed x{\sigma _{\alpha \beta }+\sigma _{\alpha \beta }}\slashed x\right]-i\delta _{ab}\frac{x^{2}\slashed xg_{}^{2}\langle
\overline{q}q\rangle ^{2}}{7776} -\delta _{ab}\frac{x^{4}\langle \overline{q}q\rangle \langle
g_{}^{2}G^{2}\rangle }{27648}+\ldots,
\end{eqnarray}
and the heavy quark propagator is given by \cite{Agaev:2020zad}:
\begin{eqnarray}\label{HeavyProp}
&&S_{Q}^{ab}(x)=i\int \frac{d^{4}k}{(2\pi )^{4}}e^{-ikx}\Bigg
\{\frac{\delta_{ab}\left( {\slashed k}+m_{Q}\right) }{k^{2}-m_{Q}^{2}}-\frac{g_{}G_{ab}^{\mu \nu}}{4}\frac{\sigma _{\mu\nu }\left( {%
\slashed k}+m_{Q}\right) +\left( {\slashed k}+m_{Q}\right) \sigma
_{\mu\nu}}{(k^{2}-m_{Q}^{2})^{2}} +\frac{g_{}^{2}G^{2}}{12}\delta _{ab}m_{Q}\frac{k^{2}+m_{Q}{\slashed k}}{%
(k^{2}-m_{Q}^{2})^{4}}\notag\\
&&+\ldots\Bigg \},
\end{eqnarray}
where 
\begin{eqnarray}\label{GluonField}
&&G_{ab}^{\mu \nu }=G_{A}^{\mu\nu
}t_{ab}^{A},\,\,~~G^{2}=G_{A}^{\mu\nu} G_{\mu \nu }^{A},
\end{eqnarray}
$A,B,C=1,\,2\,\ldots 8$; $\mu$ and $\nu$ are Lorentz indices and $t^{A}=\lambda ^{A}/2$ with
$\lambda ^{A}$ being the Gell-Mann matrices. The gluon field
strength tensor $G_{\mu\nu}^{A}\equiv G_{\mu \nu
}^{A}(0)$ is fixed at $x=0$.
Each term in quark propagator gives us an operator with the mass dimension in the  Wilson's operator product expansion (OPE). Bare-loop is perturbative term and corrections from the operators with  $d=3, \langle \overline{q}q\rangle$, $d=4, \langle G^2\rangle$, $d=5, \langle \overline{q}g_{}\sigma Gq\rangle$ and $d=6, \langle\overline{q}q\rangle^2$ are non-perturbative terms.
By inserting the heavy and light quark propagators into the correlation function we get the results including all the perturbative and non-perturbative corrections of different mass dimensions. In the calculations we consider the non-perturbative operators up to six mass dimensions.
The next step is to perform the Fourier integrals and four-integrals over momenta of the heavy quarks. 
In the calculations, as an example, there appear terms of the form:
\begin{eqnarray}\label{exampleterm}
\int d^4k\int d^4k' \int d^4x e^{i(k-p).x}\int d^4y e^{i(-k'+p').y} \frac{x_\mu y_\nu k_{\mu'} k'_{\nu'}}{(k^2-m_b^2)^l(k'^2-m_c^2)^m[(y-x)^2]^n}.\notag\\
\end{eqnarray}
 To proceed, first we use the identity \cite{Azizi:2017ubq}:
%
\begin{eqnarray}\label{intyx}
\frac{1}{[(y-x)^2]^n}&=&\int\frac{d^Dt}{(2\pi)^D}e^{-it\cdot(y-x)}~i~(-1)^{n+1}~2^{D-2n}~\pi^{D/2}  \frac{\Gamma(D/2-n)}{\Gamma(n)}\Big(-\frac{1}{t^2}\Big)^{D/2-n},
\end{eqnarray}
and substitute  $x_{\mu}\rightarrow
i\frac{\partial}{\partial p_{\mu}}$ and $y_{\mu}\rightarrow
-i\frac{\partial}{\partial p'_{\mu}}$, which leads to:
\begin{eqnarray}\label{expyx}
\int d^Dt \int d^4k\int d^4k' \int d^4x e^{i(k-p+t).x}\int d^4y e^{i(-k'+p'-t).y}  \frac{f(k,k')}{(k^2-m_b^2)^l(k'^2-m_c^2)^m}\Big(-\frac{1}{t^2}\Big)^{D/2-n}.
\end{eqnarray}
Now we perform Fourier integrals using:
\begin{eqnarray}\label{fourier}
 \int d^4x e^{i(k-p+t).x}\int d^4ye^{i(-k'+p'-t).y}= (2\pi)^4\delta^4(k-p+t) (2\pi)^4\delta^4(-k'+p'-t).
\end{eqnarray}
In this step, the two resultant four-dimensional Dirac delta functions are used to perform integrals over four $ k $ and $ k' $. The remaining D-dimensional integral over t is evaluated by the Feynman parameterization and utilizing \cite{Azizi:2017ubq}:
%
\begin{eqnarray}\label{Int}
\int d^Dt\frac{(t^2)^{m}}{(t^2+L)^{n}}=\frac{i \pi^2
(-1)^{m-n}\Gamma(m+2)\Gamma(n-m-2)}{\Gamma(2)
\Gamma(n)[-L]^{n-m-2}}.\quad
\end{eqnarray}
The final step is to calculate the imaginary parts of the obtained results by employing the following identity \cite{Azizi:2017ubq}:
\begin{eqnarray}\label{gamma}
\Gamma[\frac{D}{2}-n](-\frac{1}{L})^{D/2-n}=\frac{(-1)^{n-1}}{(n-2)!}(-L)^{n-2}ln[-L].
\end{eqnarray}
Note that for those contributions that don't have imaginary parts we follow the standard procedure of the method and calculate the contributions directly.
Finally, the function takes the following form in terms of twenty-four different Lorentz structures:
\begin{eqnarray}\label{Structures}
&&\Pi_{\mu}^{\mathrm{QCD}}(p,p',q)=\Pi^{\mathrm{QCD}}_{\slashed{p}' \gamma_{\mu}\slashed{p}}(p^{2},p'^{2},q^{2})~\slashed{p}' \gamma_{\mu}\slashed{p}+
\Pi^{\mathrm{QCD}}_{p_{\mu} \slashed {p}'\slashed {p}}(p^{2},p'^{2},q^{2})~p_{\mu} \slashed {p}'\slashed {p}+
\Pi^{\mathrm{QCD}}_{p_{\mu}' \slashed {p}'\slashed {p}}(p^{2},p'^{2},q^{2})~p_{\mu}' \slashed {p}'\slashed {p}+\Pi^{\mathrm{QCD}}_{p'_\mu\slashed {p}'\gamma_5}(p^{2},p'^{2},q^{2})\notag\\
&&p'_\mu\slashed {p}'\gamma_5+
\Pi^{\mathrm{QCD}}_{p'_\mu\slashed {p}'\slashed{p}\gamma_5}(p^{2},p'^{2},q^{2})~p'_\mu\slashed {p}'\slashed{p}\gamma_5+
\Pi^{\mathrm{QCD}}_{\slashed {p}'\gamma_\mu\gamma_5}(p^{2},p'^{2},q^{2})~\slashed {p}'\gamma_\mu\gamma_5+
\Pi^{\mathrm{QCD}}_{\slashed {p}'\gamma_\mu\slashed {p}\gamma_5}(p^{2},p'^{2},q^{2})~\slashed {p}'\gamma_\mu\slashed {p}\gamma_5+\Pi^{\mathrm{QCD}}_{p_{\mu} \slashed {p}' \slashed{p}\gamma_{5}}(p^{2},p'^{2},q^{2})\notag\\
&&p_{\mu} \slashed {p}' \slashed{p}\gamma_{5}+
 \Pi^{\mathrm{QCD}}_{\slashed{p}' \gamma_{\mu}}(p^{2},p'^{2},q^{2})~\slashed{p}' \gamma_{\mu}+
\Pi^{\mathrm{QCD}}_{p_\mu\slashed {p}'\gamma_5}(p^{2},p'^{2},q^{2})~p_\mu\slashed {p}'\gamma_5+
\Pi^{\mathrm{QCD}}_{p'_\mu\slashed {p}'}(p^{2},p'^{2},q^{2})~p'_\mu\slashed {p}'+
\Pi^{\mathrm{QCD}}_{p_\mu\slashed {p}'}(p^{2},p'^{2},q^{2})~p_\mu\slashed {p}'+\notag\\
&&\Pi^{\mathrm{QCD}}_{\gamma_\mu\slashed {p}\gamma_5}(p^{2},p'^{2},q^{2})~\gamma_\mu\slashed {p}\gamma_5+
\Pi^{\mathrm{QCD}}_{\gamma_{\mu}}(p^{2},p'^{2},q^{2})~\gamma_{\mu}+
\Pi^{\mathrm{QCD}}_{\gamma_{\mu}\slashed {p}}(p^{2},p'^{2},q^{2})~\gamma_{\mu}\slashed {p}+
\Pi^{\mathrm{QCD}}_{ \gamma_{\mu}\gamma_{5}}(p^{2},p'^{2},q^{2}) ~\gamma_{\mu}\gamma_{5}+\Pi^{\mathrm{QCD}}_{p_\mu\slashed {p}\gamma_5}(p^{2},p'^{2},q^{2})\notag\\
&&p_\mu\slashed {p}\gamma_5+
\Pi^{\mathrm{QCD}}_{p'_\mu\slashed {p}\gamma_5}(p^{2},p'^{2},q^{2})~p'_\mu\slashed {p}\gamma_5+
\Pi^{\mathrm{QCD}}_{p'_\mu\slashed {p}}(p^{2},p'^{2},q^{2})~p'_\mu\slashed {p}+
\Pi^{\mathrm{QCD}}_{p_\mu\slashed {p}}(p^{2},p'^{2},q^{2})~p_\mu\slashed {p}+\Pi^{\mathrm{QCD}}_{p'_\mu}(p^{2},p'^{2},q^{2})~p'_\mu+\notag\\
&&
\Pi^{\mathrm{QCD}}_{p'_\mu\gamma_5}(p^{2},p'^{2},q^{2})~p'_\mu\gamma_5+
\Pi^{\mathrm{QCD}}_{p_\mu}(p^{2},p'^{2},q^{2})~p_\mu+
\Pi^{\mathrm{QCD}}_{p_\mu\gamma_5}(p^{2},p'^{2},q^{2})~p_\mu\gamma_5.
\end{eqnarray}
Here $\Pi^{\mathrm{QCD}}_i(p^{2},p'^{2},q^{2})$ ($i$ stands for different structures) are the invariant functions defined in terms of  double dispersion integrals as follows:
\begin{eqnarray}\label{PiQCD}
\Pi^{\mathrm{QCD}}_i(p^{2},p'^{2},q^{2})&=&\int_{s_{min}}^{\infty}ds
\int_{s'_{min}}^{\infty}ds'~\frac{\rho
^{\mathrm{QCD}}_i(s,s',q^{2})}{(s-p^{2})(s'-p'^{2})} +\Gamma_i(p^2,p'^2,q^2),
\end{eqnarray}
where $s_{min}=(2m_s+m_b)^{2}$, $s'_{min}=(2m_s+m_c)^{2}$ and $\rho_i^{\mathrm{QCD}}(s,s',q^{2})$ denote the spectral densities, defined by $\rho_i^{\mathrm{QCD}}(s,s',q^{2})=\frac{1}{\pi}Im\Pi^{QCD}_i(p^2,p'^2,q^2)$. Here  $\Gamma_i(p^2,p'^2,q^2)$ represent the contributions directly calculated.
Upon performing the quark-hadron duality assumption later, the upper limits of the integrals will be altered to $s_0$ and $s'_0$, which are continuum thresholds at the initial and final states, respectively. The spectral densities include two parts and can be returned as:
\begin{equation} \label{Rhoqcd}
\rho^{\mathrm{QCD}}_i(s,s',q^{2})=\rho_i ^{Pert.}(s,s',q^{2})+\sum_{n=3}^{4}\rho_{i}^{n}(s,s',q^{2}),
\end{equation} 
where $Pert.$ stands for the perturbative contribution, $n=3$  for the quark condensates and $n= 4 $ for the gluon condensates.
The fifth and sixth dimensions represent the mixed condensates and are denoted by $\Gamma_i(p^2,p'^2,q^2)$ in Eq.(\ref{PiQCD}).
%
Now we apply the double Borel transformation   to the QCD side using \cite{Aliev:2006gk}:



\begin{eqnarray}\label{BorelQCD2}
\mathbf{\widehat{B}}\frac{1}{(p^{2}-s)^m} \frac{1}{(p'^{2}-s')^n}\longrightarrow (-1)^{m+n}\frac{1}{\Gamma[m]\Gamma[n]} \frac{1}{(M^2)^{m-1}}\frac{1}{(M'^2)^{n-1}}e^{-s/M^2} e^{-s'/M'^2}.
\end{eqnarray}
As mentioned before,  the Borel transformation subtracts  the contributions of the higher resonances and continuum and enhances the ground states contributions at the initial and final channels. We also perform continuum subtraction supplied by the quark hadron assumption. As a result,  we get:
\begin{eqnarray}\label{qcd part2}
&&\Pi^{\mathrm{QCD}}_i (M^2,M'^2,s_0,s'_0,q^2)=\int _{s_{min}}^{s_0} ds\int _{s'_{min}}^{s'_0}ds' e^{-s/M^2} e^{-s'/M'^2}\rho
^{\mathrm{QCD}}_{i}(s,s',q^{2})+\mathbf{\widehat{B}}\Big[\Gamma_i(p^{2},p'^{2},q^{2})\Big],\nonumber\\
\end{eqnarray}
where  the components of  $\rho_{i}(s,s^{\prime},q^2)$ and $\Gamma_i(p^{2},p'^{2},q^{2})$ are given, as an example for the structure $\slashed{p}' \gamma_{\mu}\slashed{p}$, in Appendix C.

At the last step, we match the coefficients of different structures from the phenomenological  and QCD sides to get the sum rules for the form factors in terms of QCD parameters like quark masses, quark and gluon condensates, strong coupling constant, etc. as well as the auxiliary parameters $M^2$, $M'^2$, $s_0$, $s'_0$ and $\beta$.

\section{Numerical Analysis of the form factors}\label{Sec3}

After obtaining sum rules for the form factors we analyze the results for the form factors and obtain their behavior in terms of $q^2$  which are necessary to find the widths of the weak transitions under study. 
 To this end,  we need some input parameters presented in 
Table\ \ref{inputParameter}.
\begin{table}[h!]
\caption{Input parameters used in calculations.}\label{inputParameter}
\begin{tabular}{|c|c|}
\hline 
Parameters                                             &  Values  \\
\hline 
$ m_b$                                                 & $(4.18^{+0.03}_{-0.02})~ \mathrm{GeV}$ \cite{ParticleDataGroup:2020ssz}\\
$ m_c$                                                 & $(1.27\pm0.02)~ \mathrm{GeV}$ \cite{ParticleDataGroup:2020ssz}\\
$m_s$                                                  & $(93.4^{+8.6}_{-3.4})~ \mathrm{MeV}$ \cite{ParticleDataGroup:2020ssz}\\
$ m_e $                                                & $ 0.51~~\mathrm{MeV}$ \cite{ParticleDataGroup:2020ssz}\\
$ m_\mu $                                              & $ 105~~\mathrm{MeV}$ \cite{ParticleDataGroup:2020ssz}\\
$ m_\tau $                                             & $ 1.776~~\mathrm{GeV}$ \cite{ParticleDataGroup:2020ssz}\\
$ m_{\Omega_b}$                                       & $ (6.045\pm0.012) \mathrm{GeV}$ \cite{ParticleDataGroup:2020ssz}\\
$ m_{\Omega_c} $                                      & $ (2.695\pm0.017)\mathrm{GeV}$  \cite{ParticleDataGroup:2020ssz} \\
$ G_{F} $                                              & $ 1.17\times 10^{-5} \mathrm{GeV^{-2}}$ \cite{ParticleDataGroup:2020ssz}\\
$ V_{cb} $                                             & $ (39\pm1.1)\times 10^{-3}$  \cite{ParticleDataGroup:2020ssz}\\
$ m^2_0 $                                              & $ (0.8\pm0.2) \mathrm{GeV^2}$\cite{Belyaev:1982sa,Belyaev:1982cd,Ioffe:2005ym} \\
$\tau_{\Omega_b} $                                    & $ 1.64^{+0.18}_{-0.17}\times 10^{-12} s$ \cite{ParticleDataGroup:2020ssz}\\
$\langle \bar{u} u\rangle$         & $-(0.24\pm0.01)^3 \mathrm{GeV^3}$ \cite{Belyaev:1982sa,Belyaev:1982cd} \\
$\langle \bar{s} s\rangle$           & $(0.8\pm0.1) \langle \bar{u}u\rangle \mathrm{GeV^3}$ \cite{Belyaev:1982sa,Belyaev:1982cd} \\
$\langle0|\frac{1}{\pi}\alpha_s G^2|0\rangle$          &$ (0.012\pm0.004)\mathrm{GeV^4}$ \cite{Belyaev:1982sa,Ioffe:2005ym,Belyaev:1982cd}\\
$\lambda_{\Omega_b}$                               &$0.121\pm0.012  \mathrm{GeV^3}$ \cite{Agaev:2017jyt}\\
$\lambda_{\Omega_c}$                               & $0.062\pm0.018  \mathrm{GeV^3}$ \cite{Agaev:2017jyt}\\
\hline
\end{tabular}
\end{table}
The sum rules for the form factors include some auxiliary parameters as well: The Borel parameters $M^2$ and $M'^2$,  the continuum thresholds $s_0$ and $s'_0$ and the mixing parameter $\beta$ entering the currents. We need to find the working regions for these helping parameters considering the standard requirements of the method. These conditions are pole dominance  at the initial and final channels, convergence of the OPE and relatively weak dependence of the physical quantities on the auxiliary parameters.
In technique language, to find the upper limits of Borel parameters $M^2$ and $M'^2$ we demand the pole contribution (PC) to exceed the contributions of higher states and continuum, i.e.
\begin{equation} \label{PC}
PC=\frac{\Pi^{QCD}(M^2,M'^2,s_0,s'_0)}{\Pi^{QCD}(M^2,M'^2,{\infty},{\infty})}\geq\frac{1}{2}.
\end{equation}
Their lower limits are set by the convergence of the OPE series. We require that the higher dimensional non-perturbative operator has maximally a few percent contribution. In other words we impose the condition:

\begin{equation} \label{PC2}
R(M^2, M'^2)=\frac{\Pi^{^{QCD}-dim6}(M^2,M'^2,s_0,s'_0)}{\Pi^{QCD}(M^2,M'^2,s_0,s'_0)}\leq0.05.
\end{equation}
These requirements lead to following regions for the Borel parameters:

\begin{eqnarray}
 &&9~\mathrm{GeV^2}\leq M^2 \leq 12~\mathrm{GeV^2}, \notag\\
\mbox{and} \notag\\
&&6~\mathrm{GeV^2} \leq M'^2 \leq 9~\mathrm{GeV^2}.
\end{eqnarray}
The parameters $s_{0}$  and $s'_0$ which represent the continuum thresholds  in  $\Omega_b$ and $\Omega_c$ channels, respectively correspond to the energy of the first excited states in the initial and final channels. These thresholds are determined based on the conditions ensuring the sum rules exhibit optimal stability in the allowed $M^2$ and $M'^2$ regions.
We choose the regions:
\begin{eqnarray}
&&(m_{\Omega_b}+0.1)^2~ \mathrm{GeV^2} \leq s_{0} \leq (m_{\Omega_b}+0.5)^2~ \mathrm{GeV^2},\notag\\
\mbox{and} \notag\\
&&(m_{\Omega_c}+0.1)^2~\mathrm{GeV^2}\leq s'_{0} \leq (m_{\Omega_c}+0.5)^2~ \mathrm{GeV^2},
\end{eqnarray}
which lead to a good OPE convergence and pole dominance.

\begin{figure}[h!] 
\includegraphics[totalheight=5.8cm,width=5.9cm]{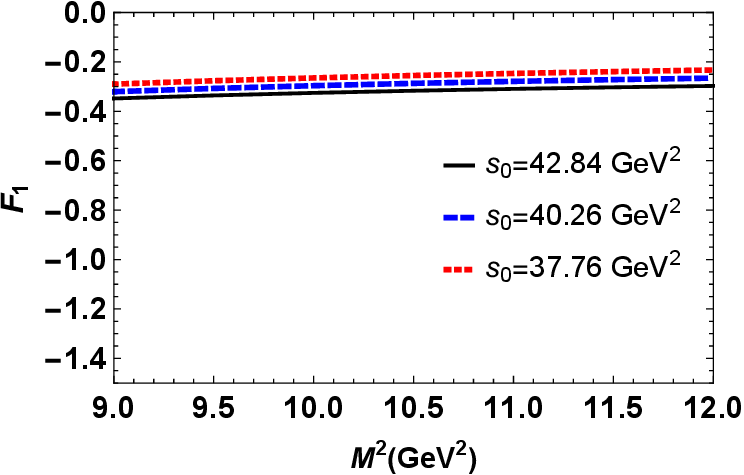}
\includegraphics[totalheight=5.8cm,width=5.9cm]{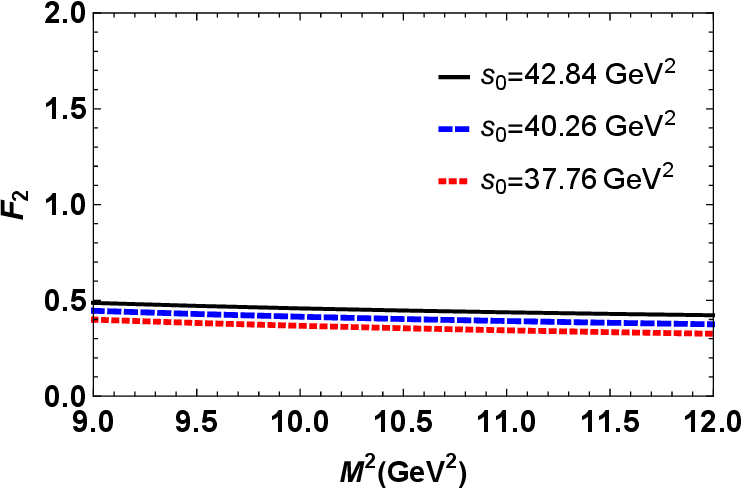}
\includegraphics[totalheight=5.8cm,width=5.9cm]{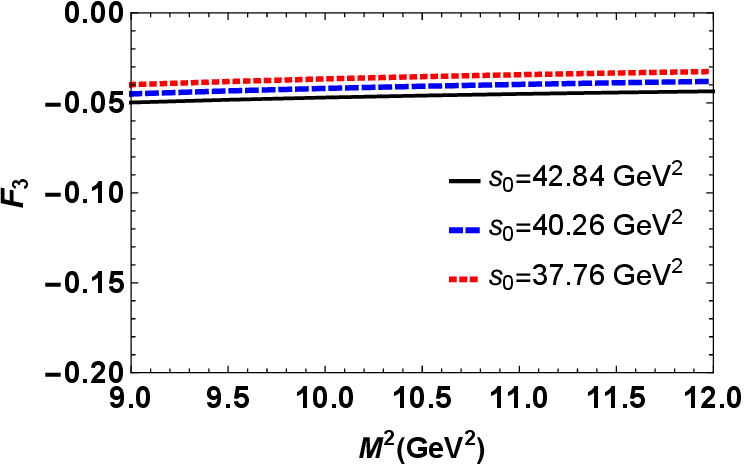}
\includegraphics[totalheight=5.8cm,width=5.9cm]{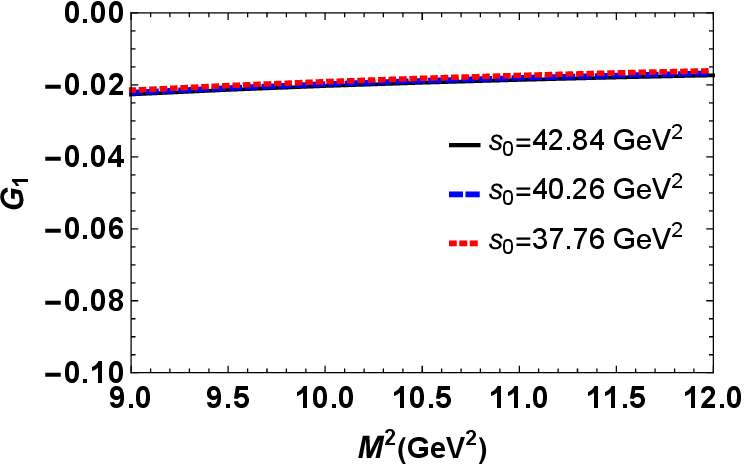}
\includegraphics[totalheight=5.8cm,width=5.9cm]{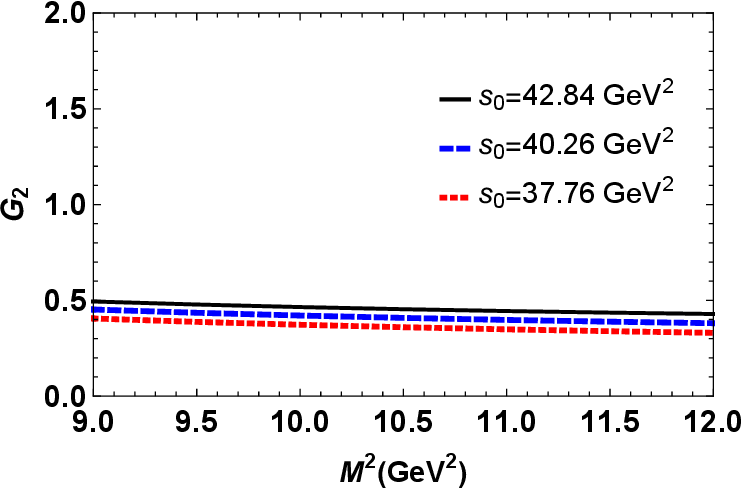}
\includegraphics[totalheight=5.8cm,width=5.9cm]{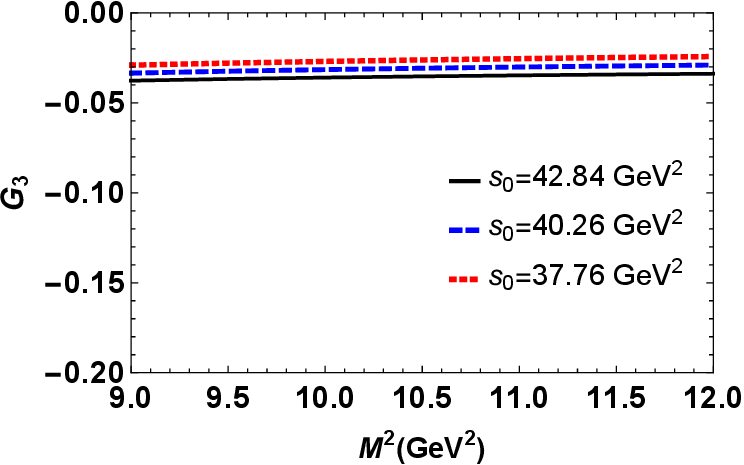}
\caption{Form factors, corresponding to the structures presented in Table \ref{Tab:parameterfit}, as functions of the Borel parameter $M^2$ at  various values 
of the parameter $s_0$,   $q^2=0$ and average values of other auxiliary parameters.}\label{Fig:BorelM}
\end{figure}
\begin{figure}[h!]
\includegraphics[totalheight=5.8cm,width=5.9cm]{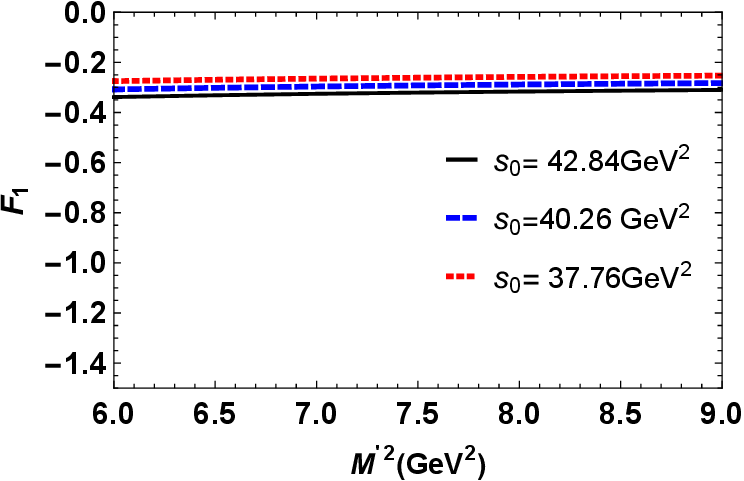}
\includegraphics[totalheight=5.8cm,width=5.9cm]{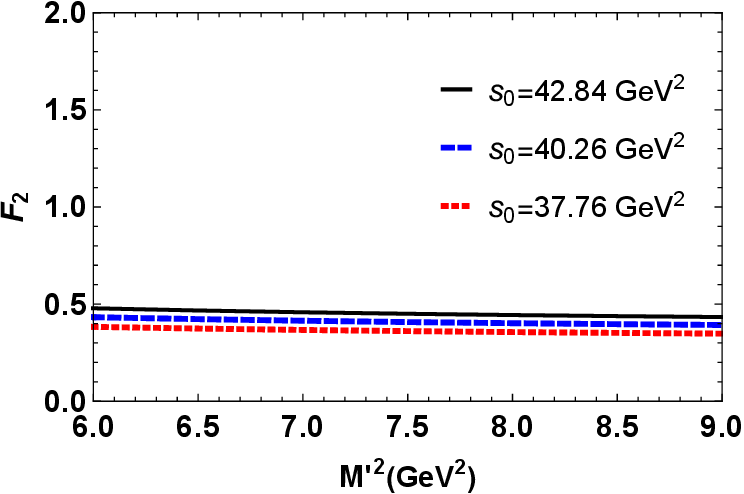}
\includegraphics[totalheight=5.8cm,width=5.9cm]{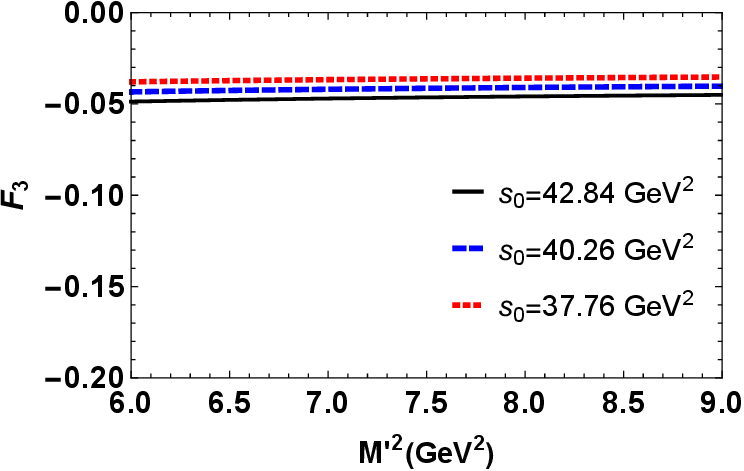}
\includegraphics[totalheight=5.8cm,width=5.9cm]{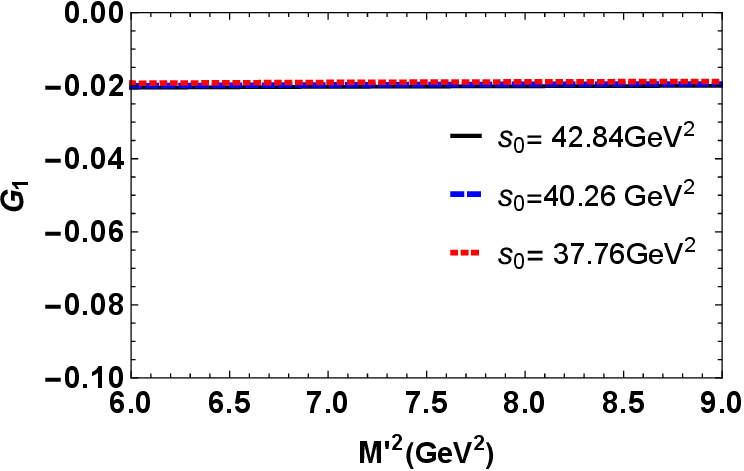}
\includegraphics[totalheight=5.8cm,width=5.9cm]{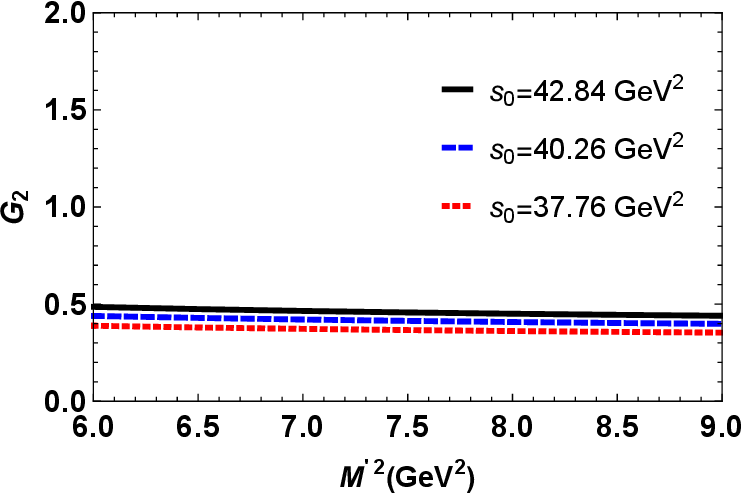}
\includegraphics[totalheight=5.8cm,width=5.9cm]{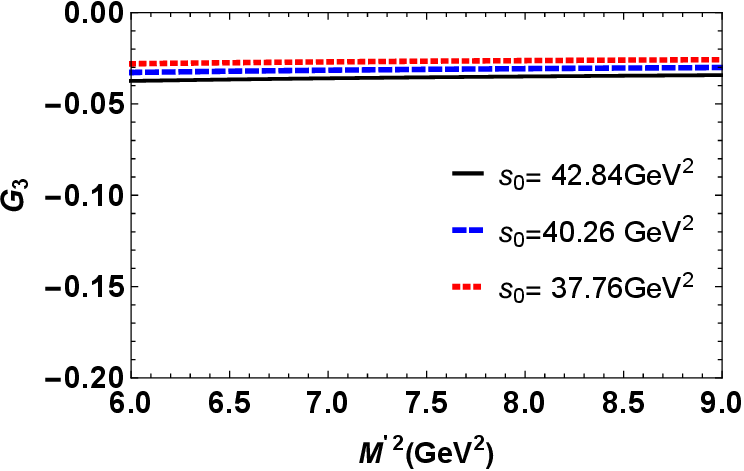}
\caption{Form factors, corresponding to the structures presented in Table \ref{Tab:parameterfit} as functions of the Borel parameter $M'^2$ at various values 
of the parameter $s_0$,  $q^2=0$ and average values of other auxiliary parameters.} \label{Fig:BorelMM}
\end{figure}
\begin{figure}[h!] 
\includegraphics[totalheight=5.8cm,width=5.9cm]{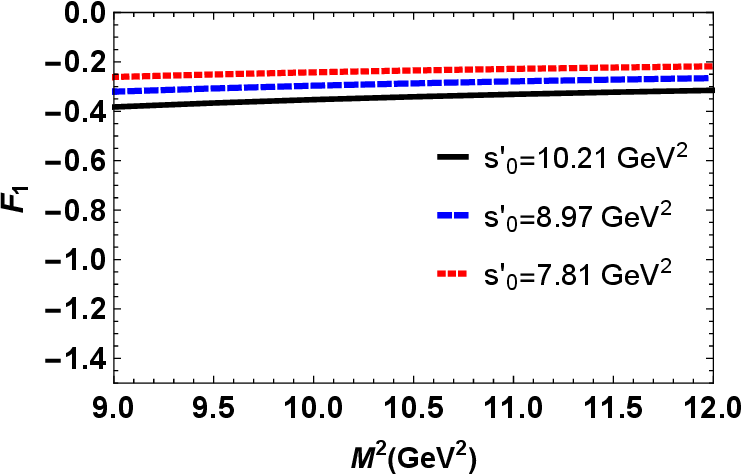}
\includegraphics[totalheight=5.8cm,width=5.9cm]{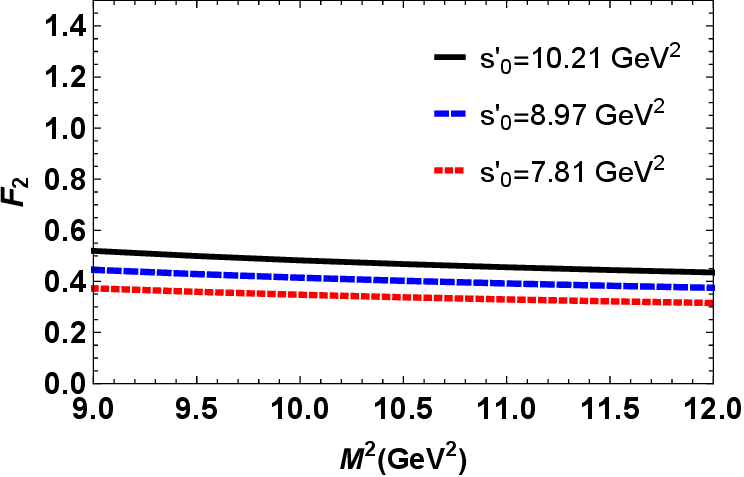}
\includegraphics[totalheight=5.8cm,width=5.9cm]{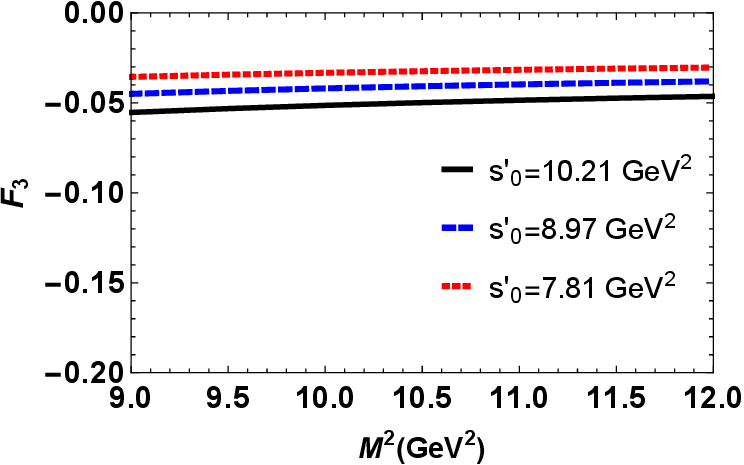}
\includegraphics[totalheight=5.8cm,width=5.9cm]{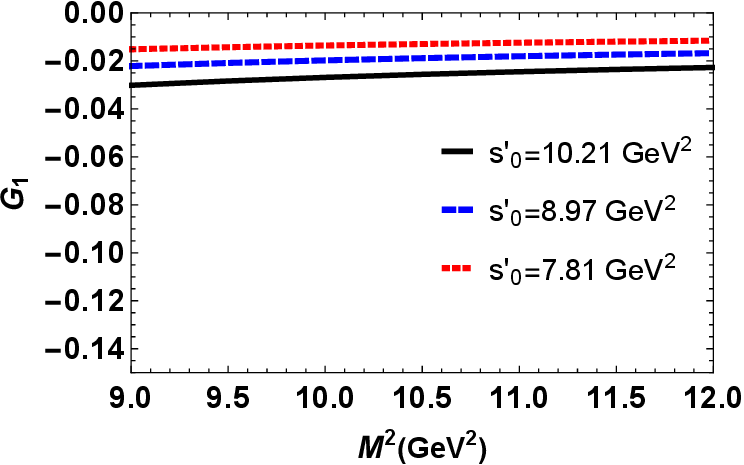}
\includegraphics[totalheight=5.8cm,width=5.9cm]{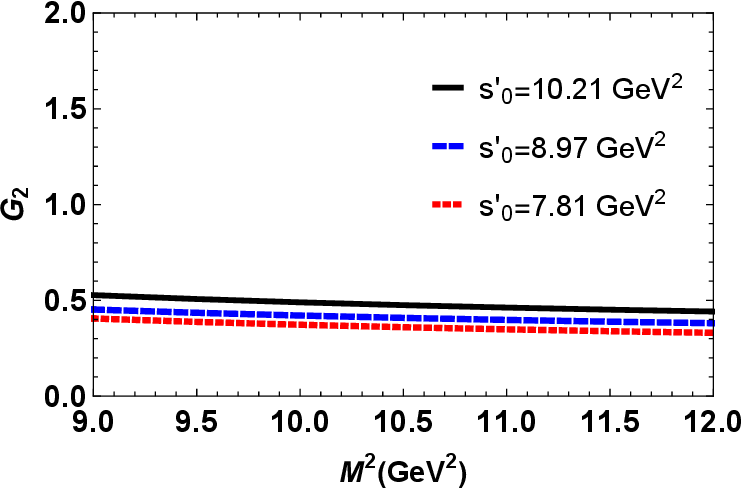}
\includegraphics[totalheight=5.8cm,width=5.9cm]{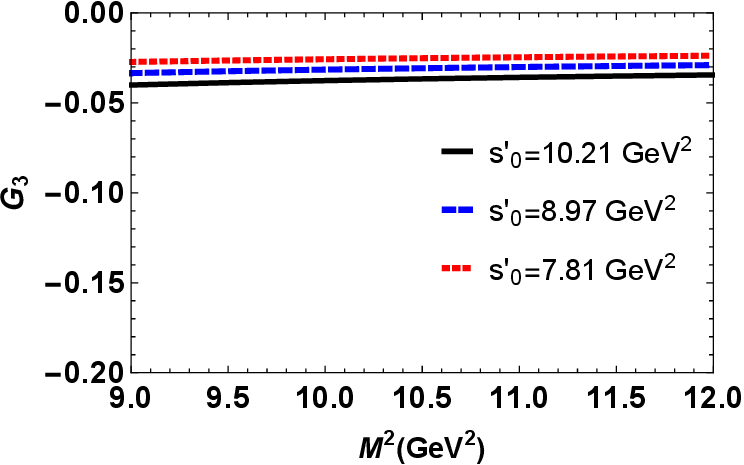}
\caption{Form factors, corresponding to the structures presented in Table \ref{Tab:parameterfit}, as functions of the Borel parameter $M^2$ at  various values 
of the parameter $s'_0$,   $q^2=0$ and average values of other auxiliary parameters.}\label{Fig:BorelMs'}
\end{figure}
\begin{figure}[h!]
\includegraphics[totalheight=5.9cm,width=5.9cm]{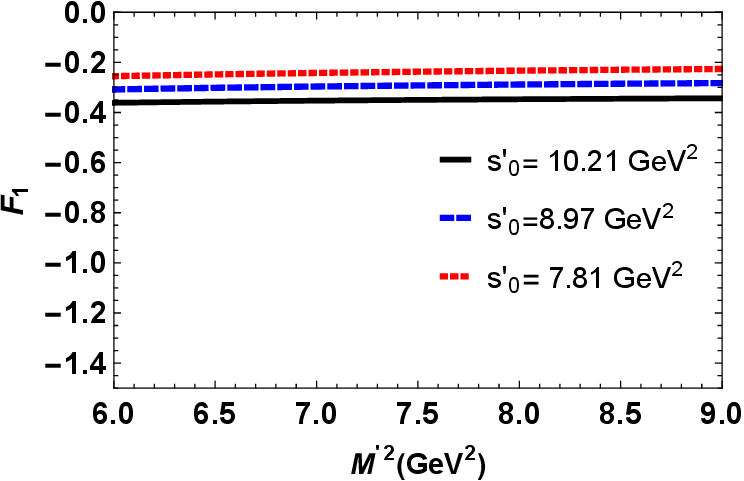}
\includegraphics[totalheight=5.9cm,width=5.9cm]{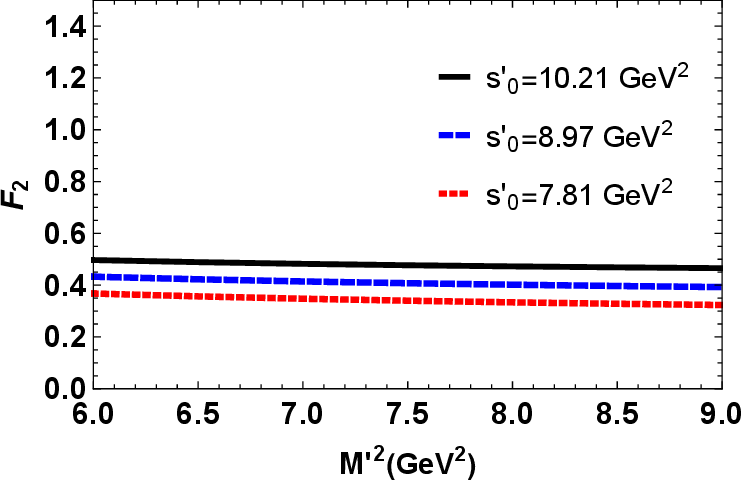}
\includegraphics[totalheight=5.9cm,width=5.9cm]{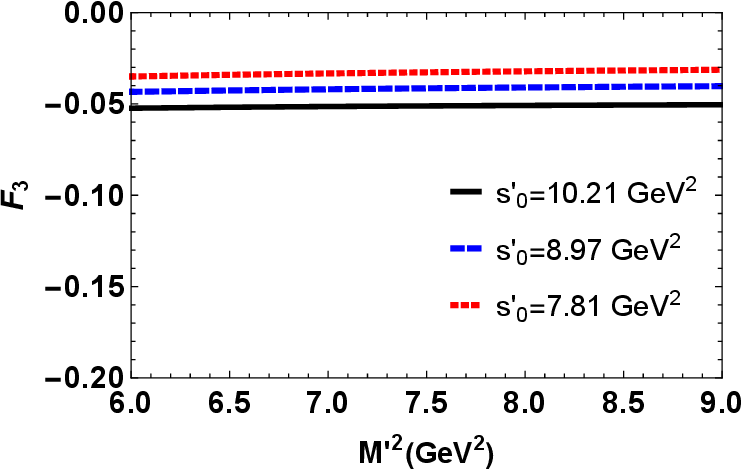}
\includegraphics[totalheight=5.9cm,width=5.9cm]{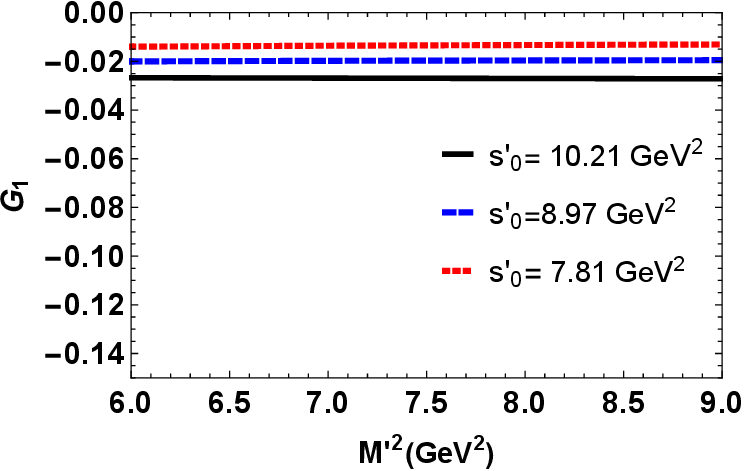}
\includegraphics[totalheight=5.9cm,width=5.9cm]{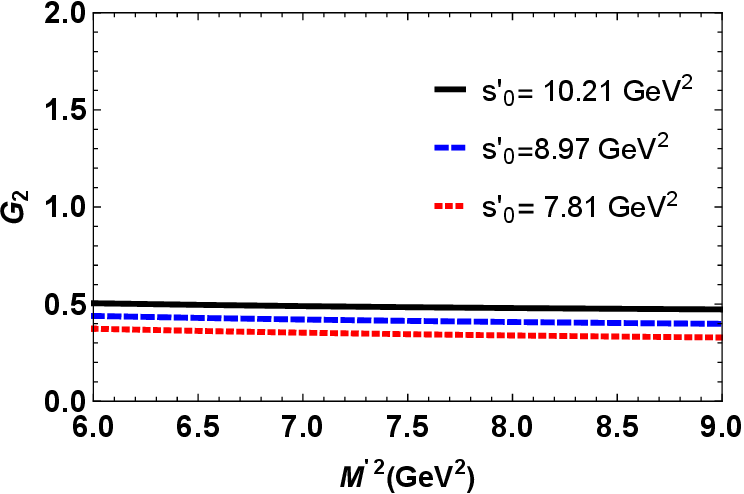}
\includegraphics[totalheight=5.9cm,width=5.9cm]{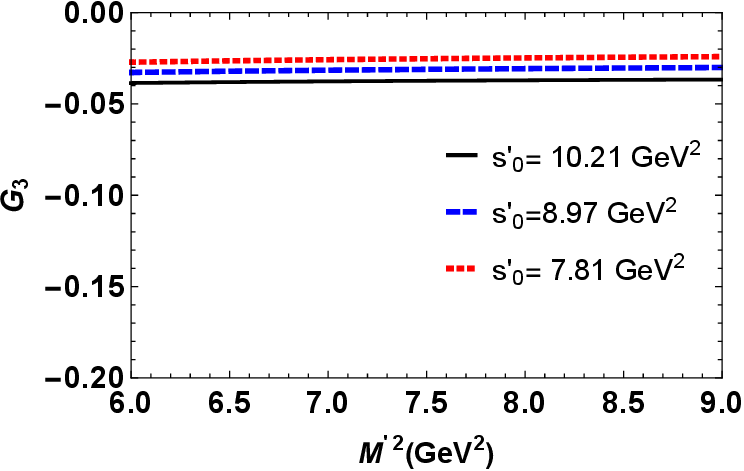}
\caption{Form factors, corresponding to the structures presented in Table \ref{Tab:parameterfit} as functions of the Borel parameter $M'^2$ at various values 
of the parameter $s'_0$,  $q^2=0$ and average values of other auxiliary parameters.} \label{Fig:BorelMMs'}
\end{figure}

\begin{figure}[h!]
\includegraphics[totalheight=6cm,width=8cm]{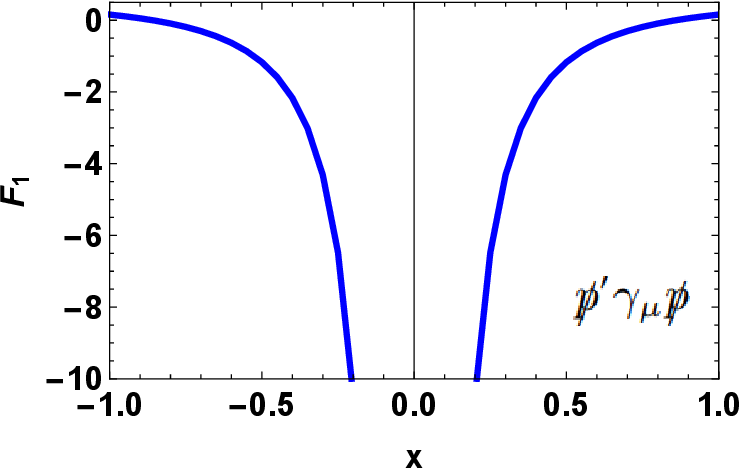}
\includegraphics[totalheight=6cm,width=8cm]{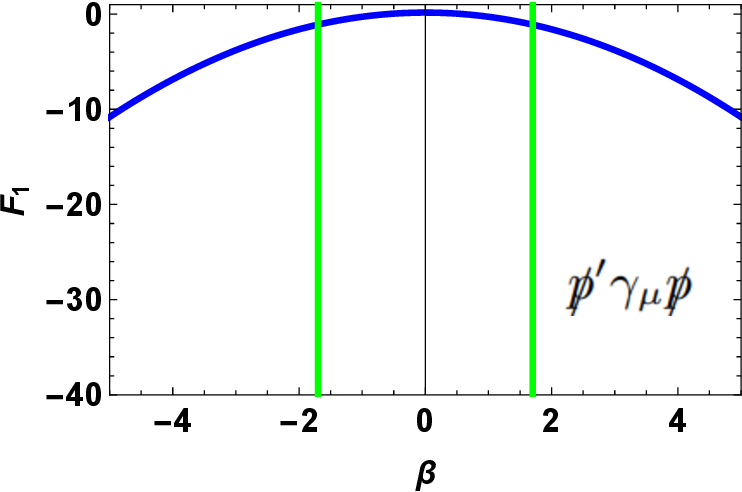}
\caption{Variations of $F_1$ form factor corresponding to the structure $\slashed {p}'\gamma_{\mu} \slashed{p}$ with respect to  $x$ and $\beta$ at  the average values of  $s_0$, $s'_0$, $M^2$ and $M'^2$ and at $q^2=0$.} \label{Fig:f1x}
\end{figure}

As is seen from Figs. \ref{Fig:BorelM},  \ref{Fig:BorelMM}, \ref{Fig:BorelMs'} and \ref{Fig:BorelMMs'} the form factors show good stability with respect to variations of $M^2$, $M'^2$, $s_0$, and $s'_0$ in their working windows. As previously mentioned, in addition to the Borel parameters and continuum thresholds we have $\beta$ parameter arising from the interpolating currents. This parameter can span the entire region from $-\infty$ to $\infty$. To confine it in a manageable range, we define  $x = cos\theta$ with $\theta = tan^{-1}\beta$ ensuring $x$ operates within the $[-1, 1]$ region. We select the region that maintains the form factors possibly unchanged. As an example, in Fig. \ref{Fig:f1x}, we depict the variations of the form factor $F_1$ with respect to $x$. From this figure we restrict this parameter as:
\begin{eqnarray}
&&-1.0\leq x \leq -0.5,\notag\\
\mbox{and} \notag \\
&&0.5\leq x \leq 1.0,
\end{eqnarray}
which are valid for all the form factors.  This is equivalent to the interval $ \beta\in [-1.73, 1.73] $, which the form factor show minimal dependency on this parameter as is seen from the plot of the form factor $F_1$ with respect to $\beta$ again in Fig. \ref{Fig:f1x}.   We shall note that the Ioffe current corresponding to $x = -0.71$ or $ \beta=-1 $ falls within the obtained region on the negative side.

After determining the working regions of the auxiliary parameters, we will analyze the behavior of the form factors in terms of
$q^2$. Our analysis shows that the form factors are well fitted to the function:
\begin{equation} \label{fitffunction}
{\cal F}(q^2)=\frac{{\cal
F}(0)}{\displaystyle\left(1-a_1\frac{q^2}{m^2_{\Omega_b}}+a_2
\frac{q^4}{m_{\Omega_b}^4}+a_3\frac{q^6}{m_{\Omega_b}^6}+a_4\frac{q^8}{m_{\Omega_b}^8}\right)}.
\end{equation}
The values of the parameters, ${\cal F}(0)$, $a_1$, $a_2$, $a_3$ and $a_4$, obtained using the average values of the auxiliary parameters, are shown in Tables~\ref{Tab:parameterfit} and \ref{Tab:parameterfit2} for different structures.
\begin{table}[h!]
\caption{Parameters of the fit functions for different form factors corresponding to $\Omega_b\to\Omega_c l\bar\nu_{\ell}$ transition.}\label{Tab:parameterfit}
\begin{ruledtabular}
\begin{tabular}{|c|c|c|c|c|c|c|}
            & $F_1(q^2): \slashed {p}'\gamma_{\mu} \slashed{p} $ & $F_2(q^2):p_{\mu}\slashed {p'}\slashed {p}$  & $F_3(q^2):p'_{\mu}\slashed {p'}\slashed {p}$   & $G_1(q^2):\slashed {p}'\gamma_{\mu} \slashed{p}\gamma_5 $ & $G_2(q^2):p_{\mu}\slashed {p}'\slashed {p}\gamma_5$  & $G_3(q^2):p'_{\mu}\slashed {p}'\slashed {p}\gamma_5$       \\
\hline
${\cal F}(q^2=0)$ & $-0.28\pm0.08$        & $0.39\pm0.10$      & $-0.04\pm0.01$     & $-0.018\pm0.007$  & $0.40\pm0.11$  & $-0.037\pm0.009$  \\
$a_1$           & $1.38$          & $1.32$            & $2.16$           & $0.47$           & $1.32$            &$ 2.16$              \\
$a_2$           & $0.25$         & $0.22$           & $1.38$             & $-0.01$          & $0.22$           & $1.38$           \\
$a_3$           & $0.08$            & $0.07$           & $-0.19$          & $0.12$          & $0.072$           & $-0.19$           \\
$a_4$           & $0.01$           & $0.003$           & $-0.045$            & $-0.06$         & $0.004$          & $-0.05$           \\
\end{tabular}
\end{ruledtabular}
\end{table}
%
\begin{figure}[h!] 
\includegraphics[totalheight=5cm,width=5.8cm]{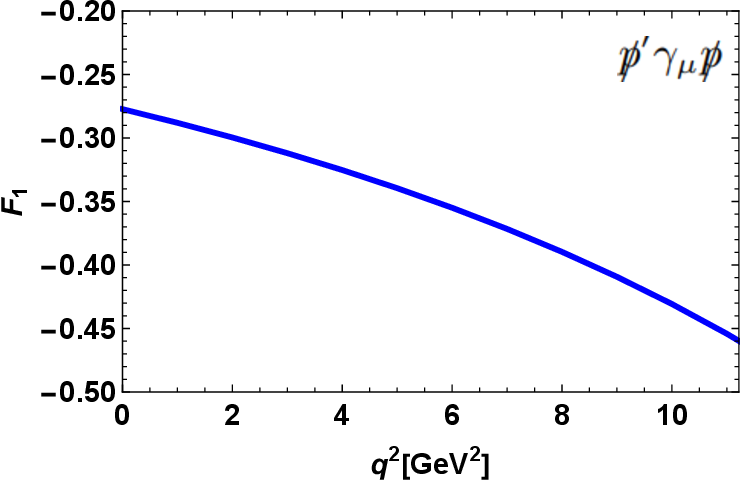}
\includegraphics[totalheight=5cm,width=5.8cm]{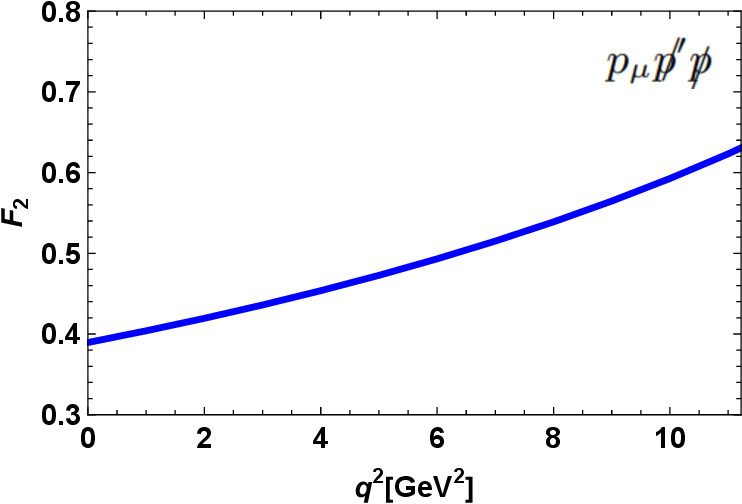}
\includegraphics[totalheight=5cm,width=5.8cm]{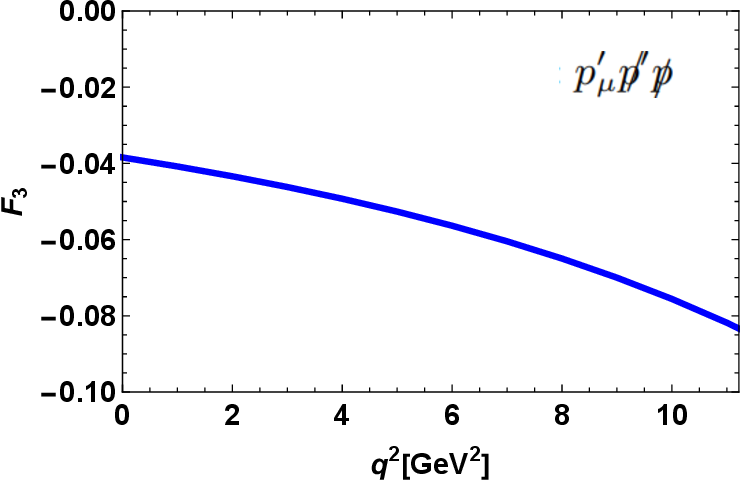}
\includegraphics[totalheight=5cm,width=5.8cm]{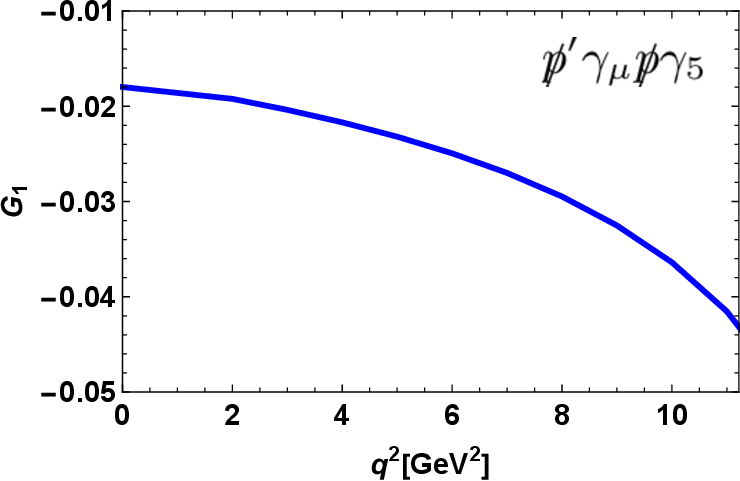}
\includegraphics[totalheight=5cm,width=5.8cm]{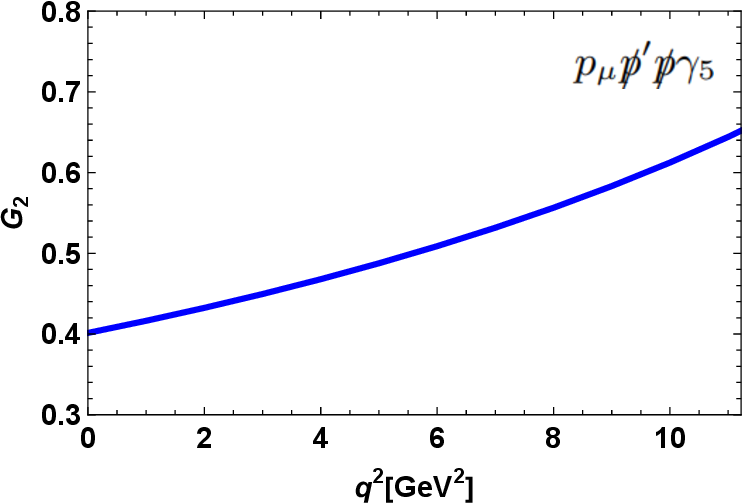}
\includegraphics[totalheight=5cm,width=5.8cm]{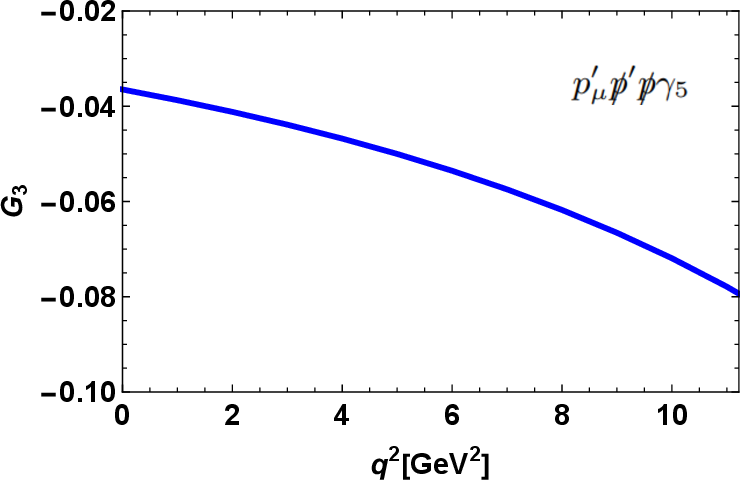}
\caption{The form factors $F_1$,  $F_2$, $F_3$,  $G_1$ , $G_2$ and $G_3$, corresponding to the structures in Table \ref{Tab:parameterfit}, as functions of $q^2$ at average values of auxiliary parameters and Ioff point.}\label{Fig:formfactor1}
\end{figure}

\begin{figure}[h!] 
\includegraphics[totalheight=5cm,width=5.8cm]{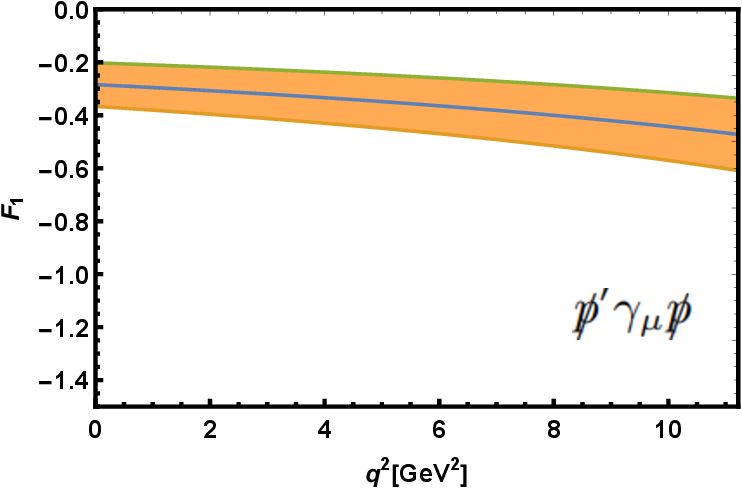}
\includegraphics[totalheight=5cm,width=5.8cm]{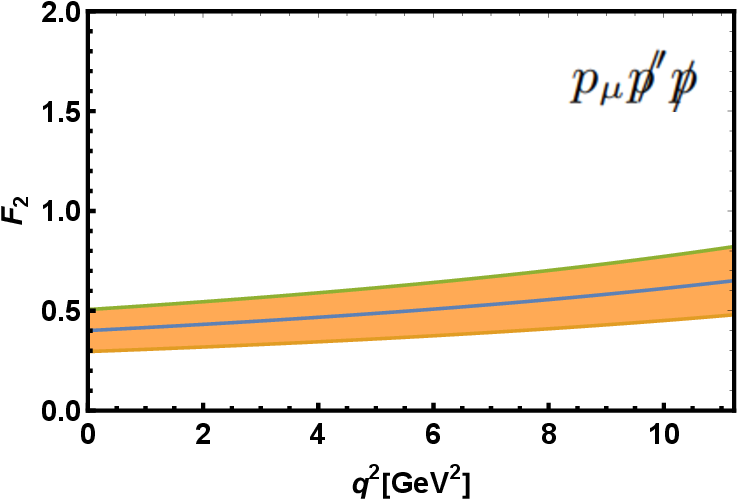}
\includegraphics[totalheight=5cm,width=5.8cm]{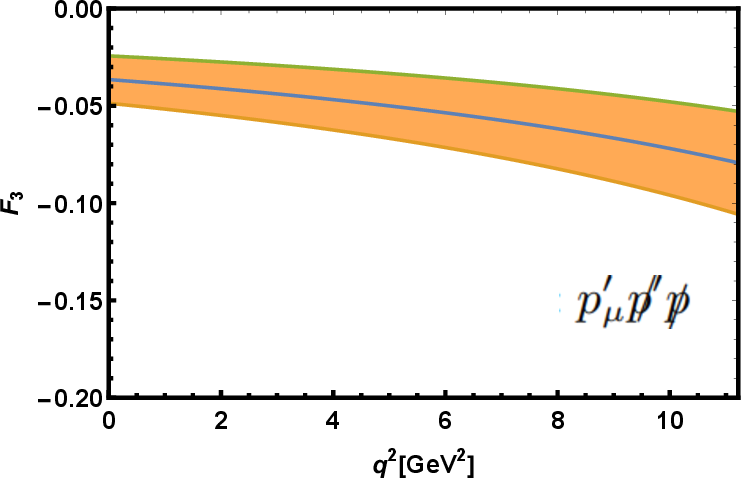}
\includegraphics[totalheight=5cm,width=5.8cm]{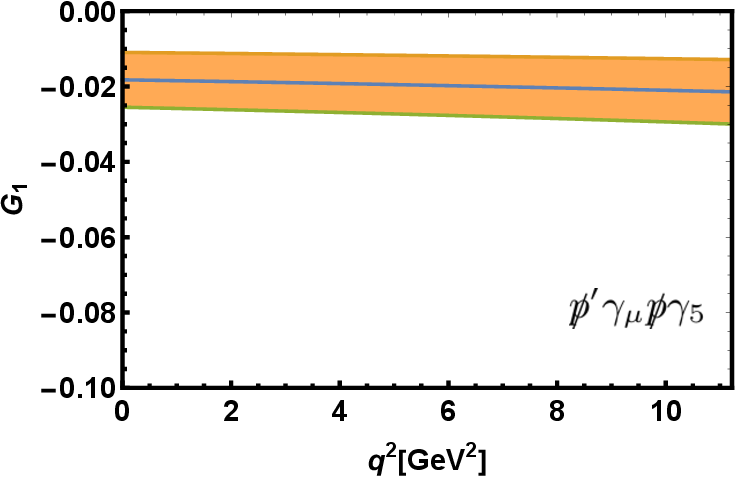}
\includegraphics[totalheight=5cm,width=5.8cm]{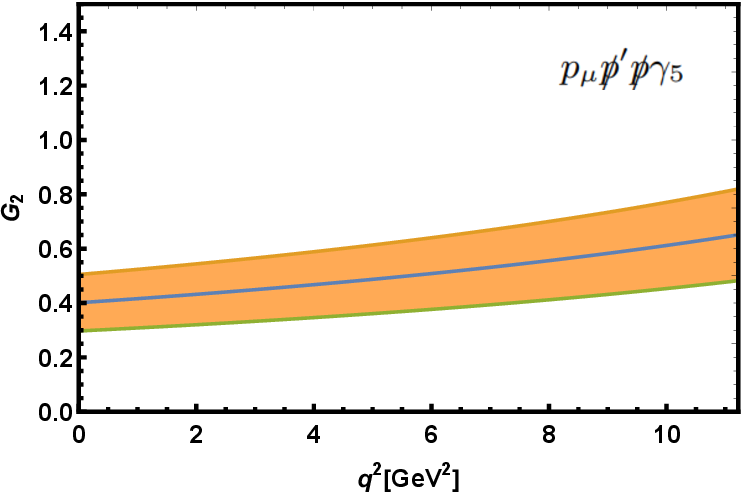}
\includegraphics[totalheight=5cm,width=5.8cm]{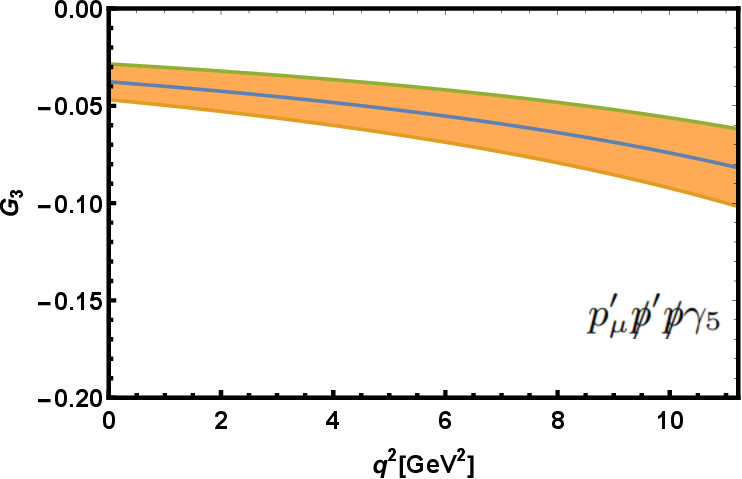}
\caption{The form factors corresponding to the structures in Table \ref{Tab:parameterfit} with their errors at the average values of auxiliary parameters.}\label{Fig:formfactorserror1}
\end{figure}

As the QCD sum rules for form factors rely on structures, multiple choices are available for each form factor. We select the best options  considering  the Borel, continuum and $x$ parameter working regions, ensuring relatively less uncertainties of the results. Generally, structures with more momenta lead to more stability. The selected structures for the form factors are shown in Table \ref{Tab:parameterfit}.
 $F1$, $G1$, and $G2$ are determined based on  fixed structures as shown in Table \ref{Tab:parameterfit}, however, for $F2$, $F3$, and $G3$, we have two more alternative structure selections that are seen in Table \ref{Tab:parameterfit2}. The presented uncertainties for the form factors at $q^2=0$ in Tables \ref{Tab:parameterfit} and \ref{Tab:parameterfit2} are due to the uncertainties in the calculations of the working regions for the auxiliary parameters as well as errors of other input values.

\begin{figure}[h!] 
\includegraphics[totalheight=5cm,width=5.8cm]{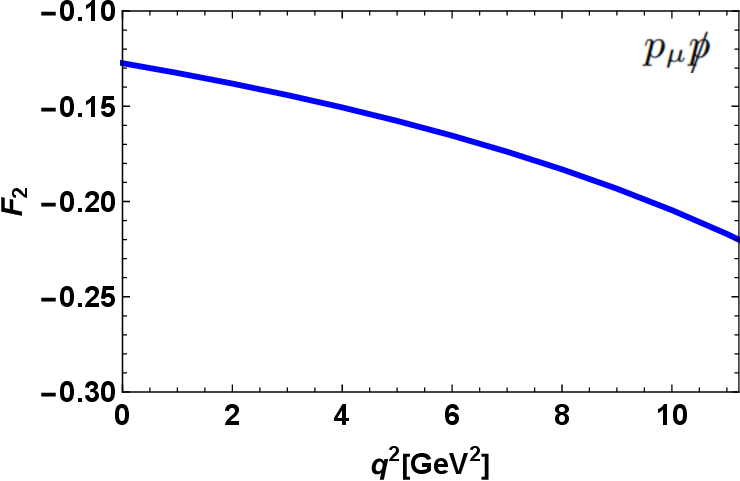}
\includegraphics[totalheight=5cm,width=5.8cm]{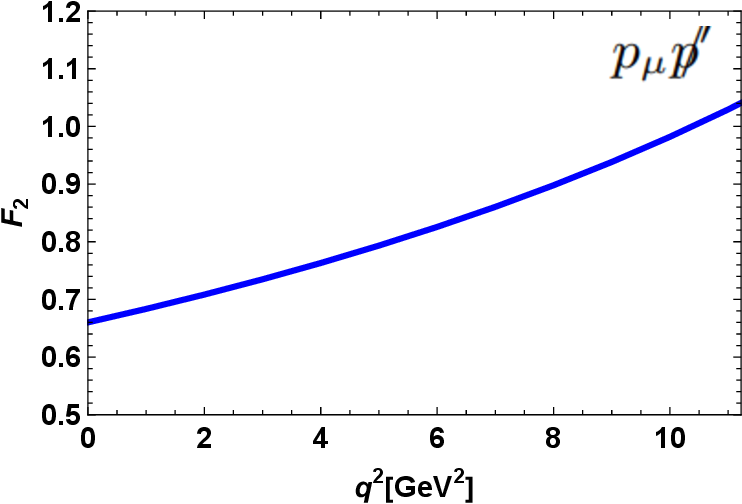}
\includegraphics[totalheight=5cm,width=5.8cm]{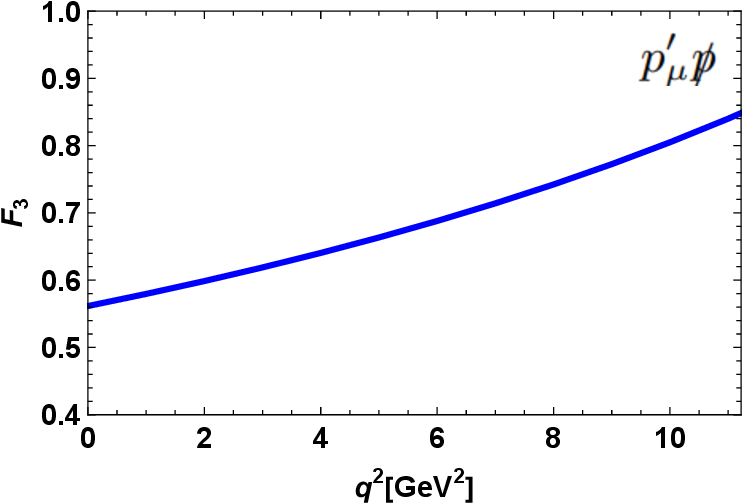}
\includegraphics[totalheight=5cm,width=5.8cm]{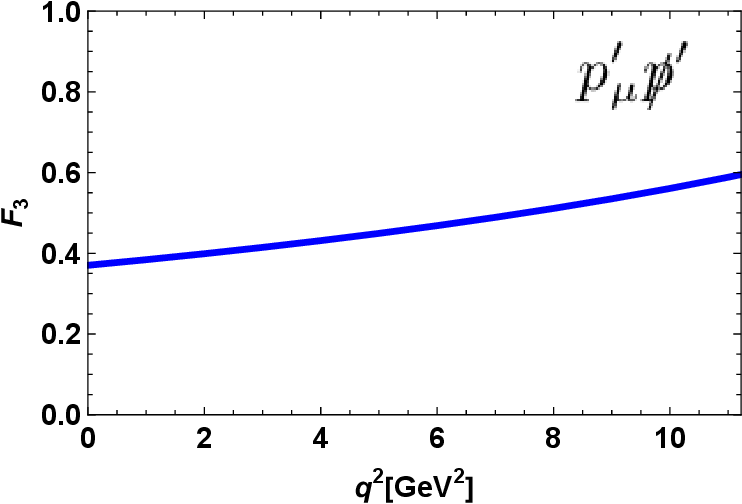}
\includegraphics[totalheight=5cm,width=5.8cm]{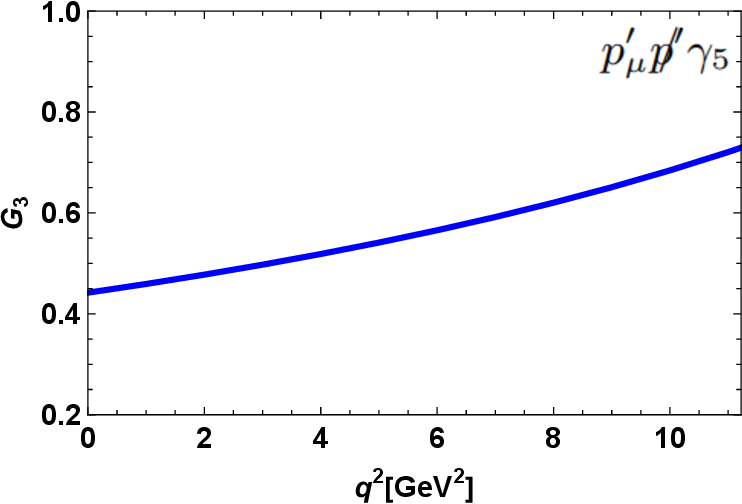}
\includegraphics[totalheight=5cm,width=5.8cm]{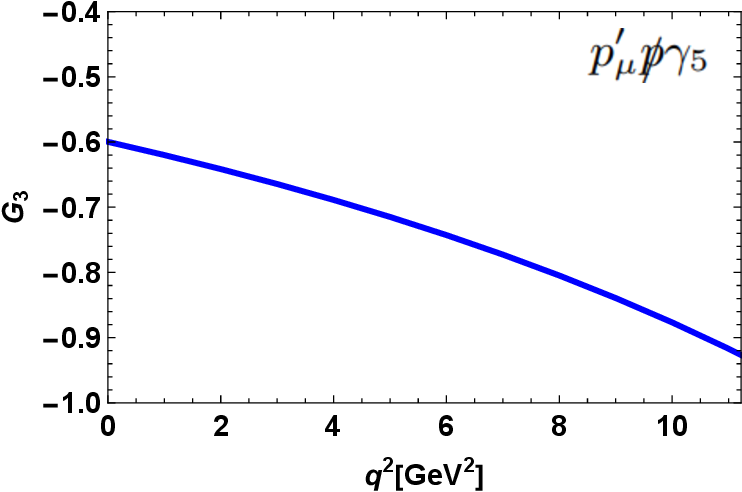}
\caption{The form factors $F_2$, $F_3$ and $G_3$ corresponding to the structures in Table \ref{Tab:parameterfit2} as functions of $q^2$ at average values of auxiliary parameters and Ioff point.}\label{Fig:formfactor2}
\end{figure}

\begin{figure}[h!] 
\includegraphics[totalheight=5cm,width=5.8cm]{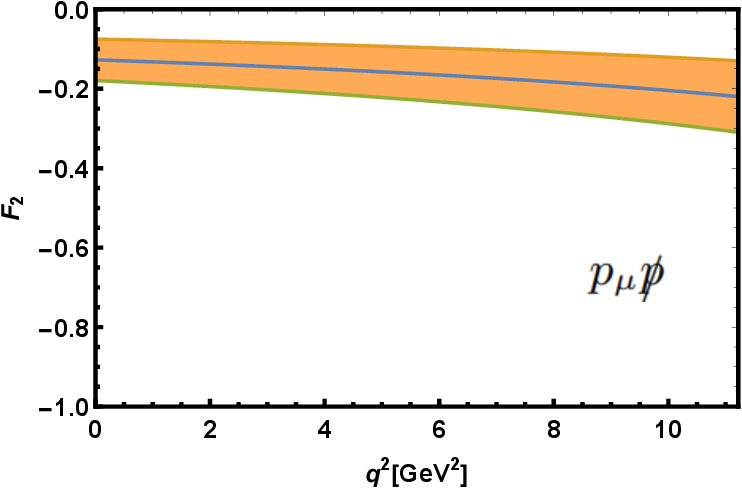}
\includegraphics[totalheight=5cm,width=5.8cm]{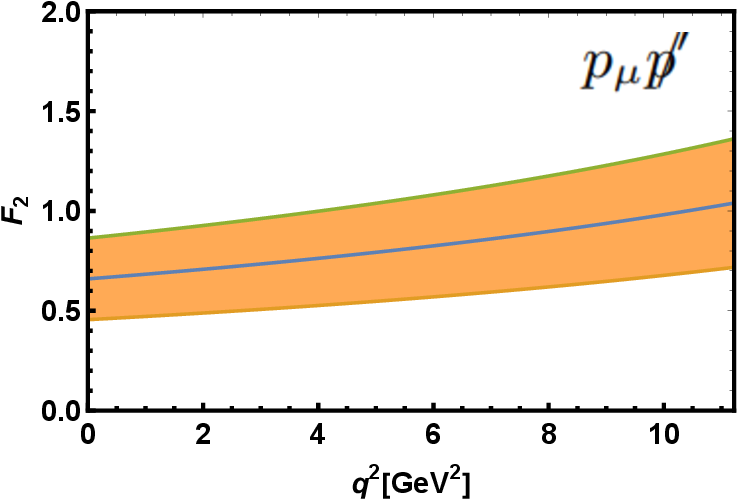}
\includegraphics[totalheight=5cm,width=5.8cm]{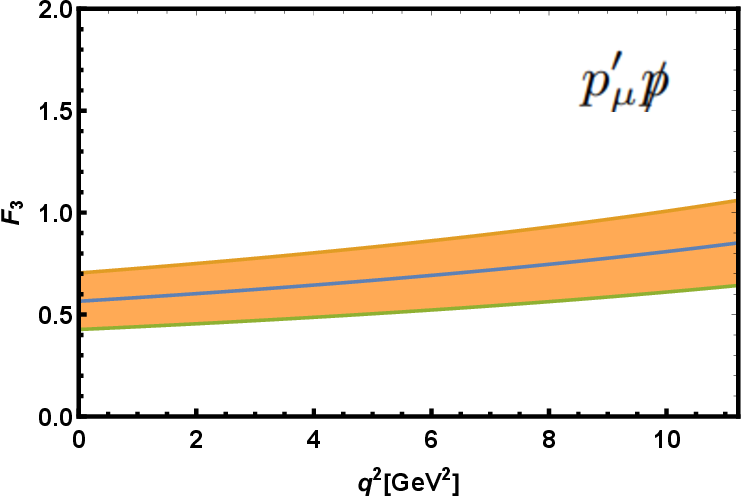}
\includegraphics[totalheight=5cm,width=5.8cm]{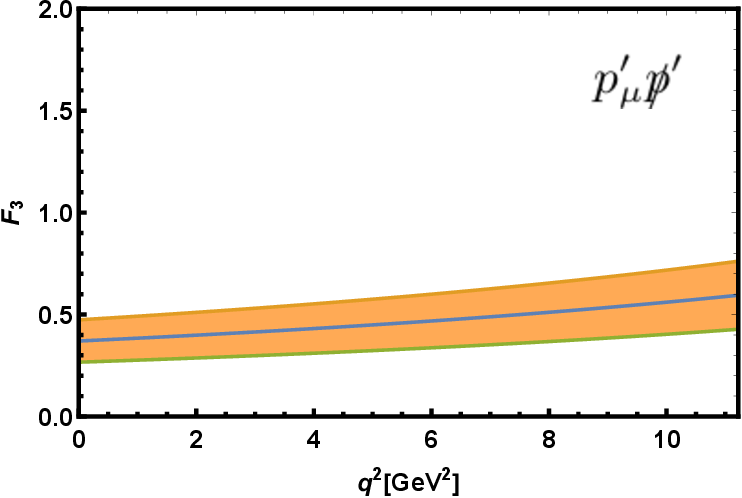}
\includegraphics[totalheight=5cm,width=5.8cm]{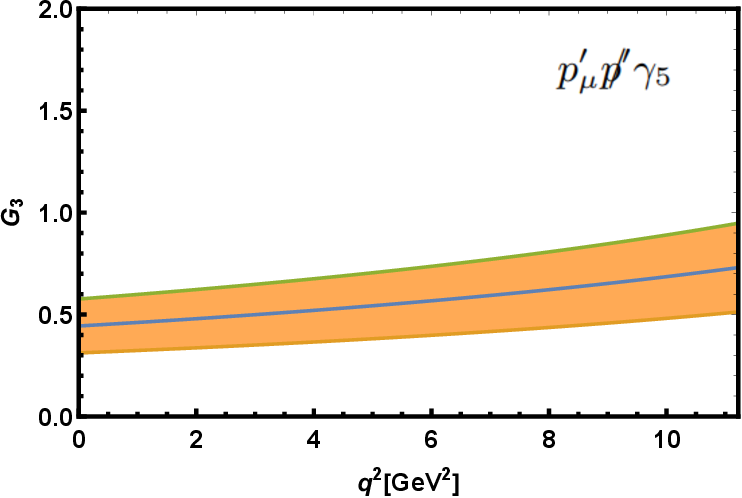}
\includegraphics[totalheight=5cm,width=5.8cm]{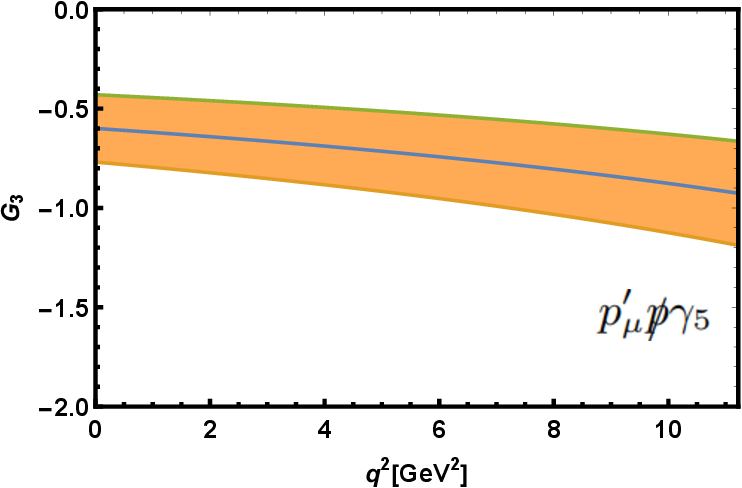}
\caption{The form factors corresponding to the structures in Table \ref{Tab:parameterfit2} with their errors at the average values of auxiliary parameters.}\label{Fig:formfactorserror2}
\end{figure}

Fig. \ref{Fig:formfactor1} shows the form factors $ F_i $ and $ G_i $ as  functions of $q^2$ at average values of $s_0,s'_0,M^2, M'^2$ and Ioffe point. Meanwhile, Fig. \ref{Fig:formfactorserror1} shows the same behavior but considering the uncertainties of the form factors. Figs. \ref{Fig:formfactor2} and \ref{Fig:formfactorserror2} depict  the behaviors of the form factors without and with uncertainties corresponding to the structures presented in Table \ref{Tab:parameterfit2}. As it is expected from weak decays,  the form factors grow with increasing the $q^2$. We will use the fit functions of the form factors in allowed physical region, $ m_l^2\leq q^2 \leq (m_{\Omega_b}-m_{\Omega_c})^2$, to find the exclusive widths and branching ratios in next section.

\begin{table}[h!]
\caption{Parameters of the fit functions for other possible structures for form factors of
$\Omega_b\to\Omega_c$ transition.}\label{Tab:parameterfit2}
\begin{ruledtabular}
\begin{tabular}{|c|c|c|c|c|c|c|}
                  & $F_2(q^2):p_{\mu}\slashed{p}$ & $F_2(q^2):p_{\mu}\slashed{p}'$  & $F_3(q^2):p'_{\mu}\slashed{p}$   &  $F_3(q^2):p'_{\mu}\slashed{p}'$   & $G_3(q^2):p'_{\mu} \slashed{p}'\gamma_5$ & $G_3(q^2):p'_{\mu}\slashed{p}\gamma_5$    \\
\hline
${\cal F}(q^2=0)$ & $-0.13\pm0.05$  & $0.66\pm0.21$    & $0.57\pm0.14$ &$0.37\pm0.12$    & $0.44\pm0.13$  & $-0.60\pm0.17$   \\
$a_1$           & $1.42$           & $1.25$                & $1.14$            &$1.33$                       & $1.38$            &$ 1.2$              \\  
$a_2$           & $0.11$           & $0.19$                & $0.13$            &$0.31$                         & $0.31$           & $0.15$           \\
$a_3$           & $0.16$           & $0.06$                & $0.05$            &$0.06$                          & $0.07$           & $0.05$           \\
$a_4$           & $0.15$           & $-0.006$             & $-0.001$        &$-0.014$                        & $-0.013$         & $-0.002$           \\    
\end{tabular}
\end{ruledtabular}
\end{table}

\section{Decay Width and Branching Ratio}~\label{Sec4}
Now, we are in a position to evaluate the decay widths and branching fractions of the
semileptonic $\Omega_b\rightarrow  \Omega_c\bar{\ell}{\nu}$ transitions in all lepton channels using the fit functions of the form factors determined in the previous section.
We utilize the following formula \cite{Faustov:2016pal,Gutsche:2014zna,Migura:2006en,Korner:1994nh,Bialas:1992ny}:

\begin{equation}\label{eq:dgamma}
\frac{d\Gamma(\Omega_b\to\Omega_c\ell\bar\nu_\ell)}{dq^2}=\frac{G_F^2}{(2\pi)^3}
|V_{cb}|^2\frac{\lambda^{1/2}(q^2-m_\ell^2)^2}{48m_{\Omega_b}^3q^2}{\cal
H}_{tot}(q^2),
\end{equation}
where $\lambda$ is defined as  $\lambda\equiv\lambda(m^2_{\Omega_b}, m^2_{\Omega_c}, q^2)=m^4_{\Omega_b}+m^4_{\Omega_c}+q^4-2(m^2_{\Omega_b}m^2_{\Omega_c}+m^2_{\Omega_b}q^2+m^2_{\Omega_c}q^2)$, $m_l$ stands for the lepton mass and ${\cal
H}_{tot}(q^2)$ refers to the total helicity and it is defined as:

\begin{equation}
 \label{eq:hh}
 {\cal H}_{tot}(q^2)=[{\cal H}_U(q^2)+{\cal H}_L(q^2)] \left(1+\frac{m_\ell^2}{2q^2}\right)+\frac{3m_\ell^2}{2q^2}{\cal H}_S(q^2).
\end{equation}

The relevant parity conserving helicity structures are expressed as:

\begin{eqnarray}
  \label{eq:hhc}
&&{\cal H}_U(q^2)=|H_{+1/2,+1}|^2+|H_{-1/2,-1}|^2,\notag\\
&&{\cal H}_L(q^2)=|H_{+1/2,0}|^2+|H_{-1/2,0}|^2,\notag\\
&&{\cal H}_S(q^2)=|H_{+1/2,t}|^2+|H_{-1/2,t}|^2,\notag\\
\end{eqnarray}
where different helicity amplitudes  are parameterized  in terms of the transition  form factors,   $F_i$ and $G_i$: 
\begin{eqnarray}
  \label{eq:ha}
  H^{V,A}_{+1/2,\, 0}&=&\frac1{\sqrt{q^2}}{\sqrt{2m_{\Omega_b}m_{\Omega_c}(\alpha\mp 1)}}
[(m_{\Omega_b} \pm m_{\Omega_c}){\cal F}^{V,A}_1(\alpha) \pm m_{\Omega_c}
(\alpha\pm 1){\cal F}^{V,A}_2(\alpha)\cr
&& \pm m_{\Omega_{b}} (\alpha\pm 1){\cal F}^{V,A}_3(\alpha)],\cr
 H^{V,A}_{+1/2,\, 1}&=&-2\sqrt{m_{\Omega_b}m_{\Omega_c}(\alpha\mp 1)}
 {\cal F}^{V,A}_1(\alpha),\cr
H^{V,A}_{+1/2,\, t}&=&\frac1{\sqrt{q^2}}{\sqrt{2m_{\Omega_b}m_{\Omega_c}(\alpha\pm 1)}}
[(m_{\Omega_b} \mp m_{\Omega_c}){\cal F}^{V,A}_1(\alpha) \pm(m_{\Omega_b}- m_{\Omega_c}\alpha
){\cal F}^{V,A}_2(\alpha)\cr
&& \pm (m_{\Omega_{b}} \alpha- m_{\Omega_c}){\cal F}^{V,A}_3(\alpha)],
\end{eqnarray}
with
$$\alpha=\frac{m_{\Omega_b}^2+m_{\Omega_{c}}^2-q^2}
{2m_{\Omega_b}m_{\Omega_{c}}}.$$ 
In the above formulas, ${\cal F}^V_i\equiv F_i$, represent the vector form factors and ${\cal F}^A_i\equiv G_i$ ($i=1,2,3$) are the axial form factors, the upper(lower)  sign corresponds to the vector(the axial vector) contributions. Here, $H^{V,A}_{h',\,h_W}$ are the helicity amplitudes for weak decays including the vector (V) and the axial vector (A) currents and their indices $(h',h_W)$ are the helicities of the final baryon and the virtual W-boson. Since, the amplitudes for negative values of the helicities are  related to $H^{V,A}_{-h',\,-h_W}=\pm H^{V,A}_{h',\,h_W}$,
the total amplitude is given by: $H_{h',\,h_W}=H^{V}_{h',\,h_W}-H^{A}_{h',\,h_W}$ for the V-A currents.

We utilize the fit functions for all the form factors,  obtained in previous section, to evaluate the decay rates for all lepton channels. The average values for the widths, along with their uncertainties, are presented in Table \ref{DECAY}. Additionally, we compare our results with other predictions existing in the literature in this table. As is seen,  our result, within the errors, is consistent with the predictions of Refs. \cite{Xu:1992hj, Cheng:1995fe, Ivanov:1996fj, Ebert:2006hm, Ebert:2008oxa, Rusetsky:1997id, Sheng:2020drc, Ivanov:1999pz, Ivanov:1998ya, Du:2011nj, Zhao:2018zcb}  for the $e$/$\mu$ channel.  The prediction of  Ref. \cite{Singleton:1990ye}  differs with others considerably. Our result is also consistent with the predictions of Refs. \cite{Xu:1992hj, Sheng:2020drc} and is  close to the prediction of Ref. \cite{Han:2020sag}  considering the presented uncertainties for the $\tau$ channel. 

We also find the branching fractions at different lepton channels. We present the obtained results in 
Table \ref{BR} and compare them with other existing predictions. Considering the uncertainties, our results are in good consistency with those  of Refs.\cite{Han:2020sag, Du:2011nj, Zhao:2018zcb} for different channels.

It is instructive to evaluate the ratio of branching fractions in $\tau$ and $e$/$\mu$ channels. We find:
\begin{eqnarray}
R_{\Omega_c}=\frac{Br[\Omega_b\rightarrow \Omega_c\tau
\overline{\nu}_\tau]}{Br[\Omega_b\rightarrow \Omega_c
(e,\mu)\overline{\nu}_{(e,\mu)}]}=0.29^{+0.06}_{-0.05}.
\end{eqnarray}
This ratio is predicted (only central values) in Refs. \cite{Sheng:2020drc, Han:2020sag} as well, which are $R_{\Omega_c}=0.37$ and $R_{\Omega_c}=0.30$,  respectively.  As is seen, our result, within the errors, is consistent  with the prediction of  Ref. \cite{ Han:2020sag}  and close to that of Ref. \cite{Sheng:2020drc}. Our result together with the presented uncertainties as SM theory prediction would be very useful for comparison with future data.

\begin{table}[bth]
\caption{Decay widths (in $\mathrm{GeV}  $) for the semileptonic $\Omega_b\to\Omega_c \ell {\overline{\nu}}_\ell$ transition at different channels.}\label{DECAY}
\begin{ruledtabular}
\begin{tabular}{|c|c|c|}
                 &$\Gamma~[\Omega_b\to\Omega_c (e,\mu) {\overline{\nu}}_{(e,\mu)}]\times10^{14}$ &  $\Gamma~[\Omega_b\to \Omega_c \tau \overline{\nu}_{\tau}]\times10^{15}$    \\
\hline
Present Work &$1.10^{+0.66}_{-0.52}$& $3.11^{+1.71}_{-1.43}$     \\
\hline
$1/m_Q$ correction HQET \cite{Xu:1992hj}  & $1.32$ &$2.1$  \\
\hline
Nonrelativistic quark model \cite{Cheng:1995fe} &$1.51$   &-  \\
\hline
Spectator quark model \cite{Singleton:1990ye} &$3.55$ &-\\ 
\hline
 Relativistic three quark model \cite{Ivanov:1996fj} & $1.23$ &-\\
\hline
Relativistic in quasipotential approach \cite {Ebert:2006hm} &$0.85$  &-\\
\hline
 Relativistic three quark model \cite{Ebert:2008oxa}&$0.85$ &-\\
\hline
Independent model\cite{Sheng:2020drc} &$0.85$ & $3.5$ \\
\hline
  Relativistic three quark model \cite{Ivanov:1999pz} &  $1.11$ &-\\
\hline
Bethe-Salpeter \cite{Ivanov:1998ya}& $1.19$ &-\\
\hline
 Covariant quasipotential approach \cite{Rusetsky:1997id} &$1.72$ &-\\
\hline
Large $N_c$ HQET\cite{Han:2020sag}&- & $4.83$ \\
\hline
Large $N_{c}$ HQET  \cite{Du:2011nj}   &$1.68$  &-   \\       
\hline
Light front \cite{Zhao:2018zcb}  &$1.14$ &-  \\
\hline
\end{tabular}
\end{ruledtabular}
\end{table}

\begin{table}[bth]
\caption{Branching ratios of the semileptonic $\Omega_b\to\Omega_c \ell {\overline{\nu}}_\ell$ transition at different channels.}\label{BR}
\begin{ruledtabular}
\begin{tabular}{|c|c|c|c|c|}
                 & Present Work &  $ N_{c}$ HQET\cite{Han:2020sag} & $ N_{c}$ HQET\cite{Du:2011nj}   &   Light front\cite{Zhao:2018zcb}  \\
\hline
$Br~[\Omega_b\to\Omega_c (e,\mu) {\overline{\nu}}_{(e,\mu)}]~(\%)$ &$2.74^{+1.64}_{-1.29}$  & -& $2.82$ & $2.72$ \\
\hline
 $Br~[\Omega_b\to \Omega_c \tau \overline{\nu}_{\tau}] (\%)$  &$0.78^{+0.43}_{-0.36}$ & $1.21$&-  &-  \\       
\end{tabular}
\end{ruledtabular}
\end{table}

\section{Conclusion}~\label{Sec5}
The exploration of deviations of experimental data  from the SM predictions on some parameters related to the semileptonic $B$ decays has prompted further researches on other b-hadrons' decay channels that may exhibit similar deviations. Such possible deviations at baryonic channels can help us explore physics BSM. In this study,  we examined the semileptonic decay $\Omega_b\to \Omega_c {\ell}\nu $ in all lepton channels.
We calculated the form factors defining these weak transitions and found their fit functions in terms of $q^2$ in the allowed physical region. We used the obtained form factors to estimate the widths, branching ratios and ratio of branching fractions at different lepton channels. We compared our results with the predictions of others studies  existing in the literature.

As we said, there is no experimental data regarding these decay channels. Thanks to the new progresses in the experiments, we hope it will be possible to study such decay modes in experiments like LHCb in near future. The future data and their comparison with our results will help us not only check the values of SM parameters but also, in the case of any inconsistency in the values of decay rates, search for new physics effects. Specifically,  our prediction on $R_{\Omega_c}$ and its comparison with future related data will be very important regarding the consistency/inconsistency between the SM theory prediction and experiment. Any deviations of  data from the SM prediction can be considered as a hint for new physics effects BSM.

\section*{ACKNOWLEDGMENTS} 
K. Azizi is thankful to Iran National Science Foundation (INSF) for the partial financial support provided under the elites Grant No. 4025036.

\appendix 

\section{The  interpolating current}
In this appendix, we construct the interpolating current for the $\Omega_Q$  baryon. According to the quark model, this  single heavy
state with $ Q= $ $ b $ or $ c $, belongs to the sextet representation meaning that the interpolating current is symmetric with respect to the exchange of the light quarks. The current should also be a color singlet. By these requirements  and considering  $J^P=\frac{1}{2}^+$ as the spin-parity of this state, the general form of the interpolating current can be written as
\begin{eqnarray}
\label{kazem}
{\cal J}^{\Omega_{Q}} \sim  \epsilon^{abc} \Big\{  \Big(
s^{aT} C \Gamma s^b \Big) \tilde{\Gamma} Q^c 
+ \Big( s^{aT} C \Gamma Q^b \Big) \tilde{\Gamma}s^c
- \Big(Q^{aT} C\Gamma s^b \Big) \tilde{\Gamma} s^c  \Big\},
\end{eqnarray}
where $ \Gamma$, $ \tilde{\Gamma}=$ $ I $, $ \gamma_5 $, $ \gamma_\mu $, $ \gamma_5 \gamma_\mu$ or $ \sigma_{\mu\nu} $ are Dirac sets. We need to determine $ \Gamma$ and $ \tilde{\Gamma}$ considering all the quantum numbers and requirements. We show that the first term does not contribute to the interpolating current.  To this end, we consider the transpose of the  term $\epsilon_{abc} s^{aT} C \Gamma s^b $, which leads to
\begin{eqnarray}
[ \epsilon_{abc} s^{aT} C \Gamma s^b]^T=- \epsilon_{abc} s^{bT}\Gamma^T C^T s^{a},
\end{eqnarray}
where we considered the the Grassmann number nature of the quark fields. Using $ C^2=-1 $ and $ C^T=C^{-1} $, we get
\begin{eqnarray}
[ \epsilon_{abc} s^{aT} C \Gamma s^b]^T= \epsilon_{abc} s^{bT}C(C\Gamma^T C^{-1}) s^{a}.
\end{eqnarray}
where
\begin{eqnarray}
&&C\Gamma^T C^{-1}=\Gamma ~~~~~ ~ \mbox {for} ~~~\Gamma=I, \gamma_5, \gamma_5\gamma_\mu, \notag\\
&&\mbox{ and}\notag\\
&&C\Gamma^T C^{-1}=-\Gamma ~~~ ~\mbox {for} ~~~ \Gamma= \gamma_\mu, \sigma_{\mu\nu},
\end{eqnarray}
Switching color dummy indicies, one obtains,
\begin{eqnarray}
&&[ \epsilon_{abc} s^{aT} C \Gamma s^b]^T=-\epsilon_{abc} s^{aT} C \Gamma s^b~~~~~ ~ \mbox {for} ~~~\Gamma=I, \gamma_5, \gamma_5\gamma_\mu, \notag\\
&&\mbox{ and}\notag\\
&&[ \epsilon_{abc} s^{aT} C \Gamma s^b]^T=\epsilon_{abc} s^{aT} C \Gamma s^b~~~~~~~ ~\mbox {for} ~~~ \Gamma= \gamma_\mu, \sigma_{\mu\nu},
\end{eqnarray}
The transpose of a one by one matrix like $  \epsilon_{abc} s^{aT} C \Gamma s^b $ should be equal to itself. Thus
\begin{eqnarray}
\epsilon_{abc} s^{aT} C \Gamma s^b=0~~~~~ ~ \mbox {for} ~~~\Gamma=I, \gamma_5, \gamma_5\gamma_\mu, 
\end{eqnarray}
The $\Omega_Q$ baryon with $J^P=\frac{1}{2}^+$ consists of two $s$ quarks and a heavy quark $b$ or $c$. The spin of the attached heavy quark to the above diquark  is $\frac{1}{2}$ and the same as the spin of the baryon.  Thus the spin of the diquark is zero . This implies $\Gamma=I, \gamma_5$. Therefore $ \Gamma $ in Eq. (\ref{kazem} ) can be either $I  $ or $ \gamma_5 $, which defines our Dirac particle of spin-1/2.  Therefore, we consider both the possibilities and write the interpolating current as 
\begin{eqnarray}
\label{kazem1}
{\cal J}^{\Omega_{Q}} \sim  \epsilon^{abc} \Big\{  \Big( s^{aT} C  Q^b \Big) \tilde{\Gamma}_1s^c+\beta \Big( s^{aT} C\gamma_5  Q^b \Big) \tilde{\Gamma}_2s^c
- \Big(Q^{aT} C s^b \Big) \tilde{\Gamma}_1 s^c  -\beta  \Big(Q^{aT} C \gamma_5 s^b \Big) \tilde{\Gamma}_2 s^c\Big\},
\end{eqnarray}
where we considered the linear combinations of the two possibilities by introducing the general mixing parameter $ \beta $. Now, we should determine $ \tilde{\Gamma}_1 $ and $  \tilde{\Gamma}_2$ . They  are determined through the Lorentz and parity considerations. As previously mentioned,  the interpolating current  for the states under study  is Lorentz scalar, so one must have both of the $ \tilde{\Gamma}_1 $ and $  \tilde{\Gamma}_2$  equal to  $I $ or $ \gamma_5 $ considering the above values for $ \Gamma $. The parity transformation leads to the results $ \tilde{\Gamma}_1 =\gamma_5$ and $ \tilde{\Gamma}_2 =I$. After normalization, we obtain
\begin{eqnarray} 
{\cal J}^{\Omega_{Q}}=\frac{-1}{2}~\epsilon_{abc}\Bigg\{\Big(s^{aT}CQ^{b}\Big)\gamma_{5}s^{c} + 
\beta\Big(s^{aT}C\gamma_{5}Q^{b}\Big)s^{c}  -\Bigg[\Big(Q^{aT}Cs^{b}\Big)\gamma_{5}s^{c}  
+\beta\Big(Q^{aT}C\gamma_{5}s^{b}\Big)s^{c}\Bigg]\Bigg\},~
\end{eqnarray}

\section{The correlation function on QCD side}
After substituting the interpolating and transition currents into Eq. (\ref {CorFunc}) and  applying Wick's theorem, the following result is obtained in coordinate space:
\begin{eqnarray} \label{32 term}
&&\Pi^{OPE}_{\mu}=i^2 \int d^4x e^{-ipx}\int d^4y e^{ip'y} \frac{1}{4} \epsilon_{a'b'c'} \epsilon_{abc}\Bigg\{\gamma_5~ S^{c'a}_s(y-x) S'_b(-x)(1-\gamma_5)\gamma_\mu S_c'(y) S^{a'c}_s(y-x)\gamma_5\notag\\
&&-Tr[S'^{a'a}_s S_c(y)\gamma_\mu(1-\gamma_5) S_b(-x)] ~\gamma_5 S^{c'c}_s(y-x)\gamma_5+
\beta S^{c'a}_s(y-x) S'^{ib}_b(-x)(1-\gamma_5)\gamma_\mu S'^{b'i}_c(y) \gamma_5 S^{a'c}_s(y-x)\gamma_5\notag\\
&&-\beta Tr[S'^{a'a}_s(y-x)\gamma_5 S^{b'i}_c(y)\gamma_\mu(1-\gamma_5) S^{ib}_b(-x)] ~S^{c'c}_s(y-x)-\gamma_5 S^{c'a}_s(y-x) S'^{ib}_b(-x)(1-\gamma_5)\gamma_\mu S'^{a'i}_c(y) S^{b'c}_s(y-x)\gamma_5\notag\\
&&+Tr[S'^{b'a}_s(y-x) S'^{a'i}_c(y)\mu(1-\gamma_5)S^{ib}_b(-x)]~\gamma_5 S^{c'c}_s(y-x)\gamma_5-\beta S^{c'a}_s(x-y) S'^{ib}_b(-x) (1-\gamma_5)\gamma_\mu S'^{a'i}_c(y) \gamma_5 S^{b'c}_s(y-x) \gamma_5\notag\\
&&+\beta Tr[S'^{b'a}_s(y-x) \gamma_5 S^{a'i}_c(y) \gamma_\mu(1-\gamma_5) S^{ib}_b(-x)]~S^{c'c}_s(y-x) \gamma_5-\beta \gamma_5 S^{c'a}_s(y-x) \gamma_5 S'^{ib}_b(-x) (1-\gamma_5)\gamma_\mu S'^{b'i}_c(x) S^{a'c}_s(y-x)\notag\\
&&+\beta Tr[\gamma_5 S'^{a'a}_s(y-x) S^{b'i}_c(x)\gamma_\mu(1-\gamma_5) S^{ib}_b(-x)]~\gamma_5 S^{c'c}_s(y-x)-\beta^2 S^{c'a}_s(y-x)\gamma_5 S'^{ib}_b(-x)(1-\gamma_5)\gamma_\mu S'^{b'i}_c(y)\gamma_5 S^{a'c}_s(y-x)\notag\\
&&+\beta^2 Tr[\gamma_5 S'^{a'a}_s(y-x)\gamma_5 S^{b'i}_c(y)\gamma_\mu(1-\gamma_5) S^{ib}_b(-x)]~ S^{c'c}_s(y-x)+\beta \gamma_5 S^{c'a}_s(y-x)\gamma_5 S'^{ib}_b(-x)(1-\gamma_5)\gamma_\mu S'^{a'i}_c(y) S^{b'c}_s(y-x)\notag\\
&&-\beta Tr[\gamma_5 S'^{b'a}_s(y-x) S^{a'i}_c(y)\gamma_\mu(1-\gamma_5) S^{ib}_b(-x)]~\gamma_5 S^{c'c}_s(y-x)+\beta^2 S^{c'a}_s(y-x) \gamma_5 S'^{ib}_b(-x)(1-\gamma_5)\gamma_\mu S'^{a'i}_c(y)\gamma_5 S^{b'c}_s(y-x)\notag\\
&&-\beta^2 Tr[\gamma_5 S'^{b'a}_s(y-x)\gamma_5 S^{a'i}_c(y)\gamma_\mu(1-\gamma_5) S^{ib}_b(-y)]~ S^{c'c}_s(y-x)-\gamma_5 S^{c'b}_s(y-x) S'^{ia}_b(-x) (1-\gamma_5)\gamma_\mu S'^{b'i}_c(y) S^{a'c}_s(y-x)\gamma_5\notag\\
&&+Tr[S'^{a'b}_s(x-y) S^{b'i}_c(y)\gamma_\mu(1-\gamma_5) S^{ia}_b(-x)]~\gamma_5 S^{c'c}_s(y-x)\gamma_5-\beta S^{c'b}_s(y-x) S'^{ia}_b(-x)(1-\gamma_5)\gamma_\mu S'^{b'i}_c(y)\gamma_5 S^{a'c}_s(y-x)\gamma_5\notag\\
&&\beta Tr[S'^{a'b}s_(x-y)\gamma_5 S^{b'i}_c(y) \gamma_\mu (1-\gamma_5)S^{ia}_b(-x)]~ S^{c'c}_s(y-x)\gamma_5+\gamma_5 S^{c'b}_s(y-x) S'^{ia}_b(-x)(1-\gamma_5)\gamma_\mu S'^{a'i}_c(y) S^{b'c}_s(y-x)\gamma_5\notag\\
&&-Tr[S'^{b'b}_s(y-x)S^{a'i}_c(y)\gamma_\mu(1-\gamma_5)S^{ia}_b(-x)]~\gamma_5 S^{c'c}_s(y-x)\gamma_5+\beta S^{c'b}_s(y-x) S'^{ia}_b(-x)(1-\gamma_5)\gamma_\mu S'^{a'i}_c(y)\gamma_5 S^{b'c}_s(y-x) \gamma_5\notag\\
&&-\beta Tr[S'^{b'b}_s(y-x)\gamma_5 S^{a'i}_c(y)\gamma_\mu(1-\gamma_5) S^{ia}_b(-x)]~S^{c'c}_b(y-x)\gamma_5+\beta \gamma_5 S^{c'b}_s(y-x) \gamma_5 S'^{ia}_b(-x) (1-\gamma_5)\gamma_\mu S'^{b'i}_c(y) S^{a'c}_s(y-x)\notag\\
&&-\beta Tr[\gamma_5 S'^{a'b}_s(y-x) S^{b'i}_c(y)\gamma_\mu(1-\gamma_5) S^{ia}_b(-x)]~ \gamma_5 S^{c'c}_s(y-x)+\beta^2 S^{c'b}_s(y-x)\gamma_5 S'^{ia}_b(-x) (1-\gamma_5)\gamma_\mu S'^{b'i}_c(y)\gamma_5 S^{a'c}_s(y-x)\notag\\
&&+\beta^2 Tr[\gamma_5 S'^{a'b}_s(y-x)\gamma_5 S^{b'i}_c(y) \gamma_\mu(1-\gamma_5) S^{ia}_b(-x)]~S^{c'c}_s(y-x)-\beta \gamma_5 S^{c'b}_s(y-x)\gamma_5 S'^{ia}_b(-x) (1-\gamma_5)\gamma_\mu S'^{a'i}_c(y) S^{b'c}_s(y-x)\notag\\
&&\beta Tr[\gamma_5 S'^{b'b}_s(y-x) S^{a'i}_c(y)\gamma_\mu(1-\gamma_5) S^{ia}_b(-x)]~\gamma_5 S^{c'c}_s(y-x)-\beta^2 S^{c'b}_s(y-x)\gamma_5 S'^{ia}_b(-x) (1-\gamma_5)\gamma_\mu S'^{a'i}_c(y) \gamma_5 S^{b'c}_s(y-x)\notag\\
&&+\beta^2 Tr[\gamma_5 S'^{b'b}_s(y-x)\gamma_5 S^{a'i}_c(y)\gamma_\mu(1-\gamma_5) S^{ia}_b(-x)]~S^{c'c}_s(y-x)
\Bigg\},
\end{eqnarray}
where $S'_q=C S^T C$.

\section{Different perturbative and non-perturbative contributions}

\renewcommand{\theequation}{\Alph{section}.\arabic{equation}} \label{sec:App}
In this appendix, the explicit forms for the components of $\rho_i(s,s',q^2)$ and $\Gamma_i(p^2,p'^2,q^2)$ for the structure $\slashed{p}' \gamma_{\mu}\slashed{p}$ are given:
\begin{eqnarray} \label{RhoPert}
&\rho^{Pert.}_{\slashed{p}' \gamma_{\mu}\slashed{p}}(s,s',q^2)=\int_{0}^{1}du \int_{0}^{1-u}dv~\frac{1}{512 \pi^4 Z^2}
\Bigg\{\Big [(-1 + u) u^2\Big (-2 s' + (-q^2 + S) u\Big) + u \Big(2 S + (s + 3 s') u (-3 + 2 u) + \notag\\
 &q^2 (-2 + 3 u)\Big) v + \Big(2 s + q^2 (4 - 5 u) u + s u (-8 + 5 u) +  s' u (-4 + 7 u)\Big) v^2 + 2 \Big(-2 s + (-q^2 + 3 s + s') u\Big) v^3 + \notag\\
 & 2 s v^4 + 12 m_s^2 Z^2 - 2 L Z (Z+1)\Big] \Big[-\beta^2 +1\Big]+\Big[12 m_s Z (m_c u + m_b v)\Big]\Big[\beta^2+1\Big]\Bigg\}\Theta[L(s,s^{\prime},q^2)],
\end{eqnarray}
%
\begin{eqnarray} \label{Rho3}
\rho^3_{\slashed{p}' \gamma_{\mu}\slashed{p}}(s,s',q^2)=0,
\end{eqnarray}
\begin{equation}\label{Rho4}
\rho^4_{\slashed{p}' \gamma_{\mu}\slashed{p}}(s,s',q^2)=0,
\end{equation}
and
\begin{eqnarray}\label{gamma5,6}
\Gamma_{\slashed{p}' \gamma_{\mu\slashed{p}}}(p^2,p'^2,q^2)=\Gamma^5_{\slashed{p}' \gamma_{\mu\slashed{p}}}(p^2,p'^2,q^2)+\Gamma^6_{\slashed{p}' \gamma_{\mu\slashed{p}}}(p^2,p'^2,q^2),
\end{eqnarray}
with
\begin{eqnarray}\label{Rho5}
&&\Gamma^5_{\slashed{p}' \gamma_{\mu\slashed{p}}}(p^2,p'^2,q^2)=
\frac{\langle \bar{s}s\rangle m_o^2 m_s}{192 \pi^2 r^2 r'^2}\Big[
(m_c m_s r + m_b m_s r')(1+\beta^2)+3 r r'(\beta^2-1)\Big],
\end{eqnarray}
and
\begin{eqnarray}\label{Rho6}
&&\Gamma^6_{\slashed{p}' \gamma_{\mu\slashed{p}}}=\frac{ \langle \bar{s}s\rangle^2 }{2592 \pi^2 r^3 r'^3}\Bigg\{\Big[27 m_c^2 m_s^2 \pi^2 r R+27 \pi^2 r r' \Big(-4 r r'+ m_s^2 (-q^2 +R')\Big)+27 m_b^2 m_s^2 \pi^2 r' R\Big]\Big[1-\beta^2\Big]+\notag\\
&&\Big[- 2 m_c m_s (g_s^2 - 54 \pi^2) r^2 r'-2 m_b m_s (g_s^2 - 54 \pi^2) r r'^2 \Big]\Big[\beta^2+1\Big]\Bigg\}.
\end{eqnarray}
In the above equations we have used the following short-hand notations:
\begin{eqnarray}\label{L}
&&L(s,s^{\prime},q^2)=- m_c^2 u + s^{\prime} u - s^{\prime}
u^2 - m_b^2 v + s v+ q^2 u v - s u v - s^{\prime} u v - s v^2,\notag
\\
&&Z=u+v-1,\notag
\\
&&S=s+s',\notag
\\
&&r=m_b^2-p^2,\notag
\\
&&r'=m_c^2-p'^2,\notag
\\
&&R=r-r',\notag
\\
&&R'=r+r',
\end{eqnarray}
and $\Theta[...]$ stands for the unit-step function.

\label{sec:Num}


\begin{thebibliography}{99}
\bibitem{BaBar:2012obs}
J.~P.~Lees \textit{et al.} [BaBar],
"Evidence for an excess of $\bar{B} \to D^{(*)} \tau^-\bar{\nu}_\tau$ decays,''
\href{https://doi.org/10.1103/PhysRevLett.109.101802}{Phys. Rev. Lett. \textbf{109}, 101802 (2012)}.
\href{https://arxiv.org/pdf/1205.5442.pdf}{[arXiv:1205.5442 [hep-ex]]}.

\bibitem{LHCb:2014vgu}
R.~Aaij \textit{et al.} [LHCb],
"Test of lepton universality using $B^{+}\rightarrow K^{+}\ell^{+}\ell^{-}$ decays,''
\href{https://doi.org/10.1103/PhysRevLett.113.151601}{Phys. Rev. Lett. \textbf{113}, 151601 (2014)}.
\href{https://arxiv.org/pdf/1406.6482.pdf}{[arXiv:1406.6482 [hep-ex]]}.

\bibitem{LHCb:2017vlu}
R.~Aaij \textit{et al.} [LHCb],
"`Measurement of the ratio of branching fractions $\mathcal{B}(B_c^+\,\to\,J/\psi\tau^+\nu_\tau)$/$\mathcal{B}(B_c^+\,\to\,J/\psi\mu^+\nu_\mu)$,''
\href{https://doi.org/10.1103/PhysRevLett.120.121801}{Phys. Rev. Lett. \textbf{120}, no.12, 121801 (2018)}.
\href{https://arxiv.org/pdf/1711.05623.pdf}{[arXiv:1711.05623 [hep-ex]]}.

\bibitem{Pervin:2005ve}
M.~Pervin, W.~Roberts and S.~Capstick,
"Semileptonic decays of heavy lambda baryons in a quark model,''
\href{https://doi.org/10.1103/PhysRevC.72.035201}{Phys. Rev. C \textbf{72}, 035201 (2005).}
\href{https://arxiv.org/pdf/nucl-th/0503030.pdf}{[arXiv:nucl-th/0503030 [nucl-th]]}.

\bibitem{Faustov:2016pal}
R.~N.~Faustov and V.~O.~Galkin,
"Semileptonic decays of $\Lambda_b$ baryons in the relativistic quark model,''
\href{https://doi.org/10.1103/PhysRevD.94.073008}{Phys. Rev. D \textbf{94}, no.7, 073008 (2016)}.
\href{https://arxiv.org/pdf/1609.00199.pdf}{[arXiv:1609.00199 [hep-ph]]}.

\bibitem{Gutsche:2014zna}
T.~Gutsche, M.~A.~Ivanov, J.~G.~K\"orner, V.~E.~Lyubovitskij and P.~Santorelli,
"Heavy-to-light semileptonic decays of $\Lambda_b$ and $\Lambda_c$ baryons in the covariant confined quark model,''
\href{https://doi.org/10.1103/PhysRevD.90.114033}{Phys. Rev. D \textbf{90}, no.11, 114033 (2014)}{[erratum: Phys. Rev. D \textbf{94}, no.5, 059902 (2016)]}.
\href{https://arxiv.org/pdf/1410.6043.pdf}{[arXiv:1410.6043 [hep-ph]]}.

\bibitem{Detmold:2015aaa}
W.~Detmold, C.~Lehner and S.~Meinel,
"$\Lambda_b \to p \ell^- \bar{\nu}_\ell$ and $\Lambda_b \to \Lambda_c \ell^- \bar{\nu}_\ell$ form factors from lattice QCD with relativistic heavy quarks,''
\href{https://doi.org/10.1103/PhysRevD.92.034503}{Phys. Rev. D \textbf{92}, no.3, 034503 (2015)}.
\href{https://arxiv.org/pdf/1503.01421.pdf}{[arXiv:1503.01421 [hep-lat]]}.

\bibitem{Gutsche:2015mxa}
T.~Gutsche, M.~A.~Ivanov, J.~G.~K\"orner, V.~E.~Lyubovitskij, P.~Santorelli and N.~Habyl,
"Semileptonic decay $\Lambda_b \to \Lambda_c + \tau^- + \bar{\nu_\tau}$ in the covariant confined quark model,''
[erratum: Phys. Rev. D \textbf{91}, no.11, 119907 (2015)]
\href{https://doi.org/10.1103/PhysRevD.91.074001}{Phys. Rev. D \textbf{91}, no.7, 074001 (2015)}.
\href{https://arxiv.org/pdf/1502.04864.pdf}{[arXiv:1502.04864 [hep-ph]]}.

\bibitem{Miao:2022bga}
Y.~Miao, H.~Deng, K.~S.~Huang, J.~Gao and Y.~L.~Shen,
"\ensuremath{\Lambda}$_{b}$ \textrightarrow \ensuremath{\Lambda}$_{c}$ form factors from QCD light-cone sum rules*,''
\href{https://doi.org/10.1088/1674-1137/ac8652}{Chin. Phys. C \textbf{46}, no.11, 113107 (2022)}.
\href{https://arxiv.org/pdf/2206.12189.pdf}{[arXiv:2206.12189 [hep-ph]]}.

\bibitem{Duan:2022uzm}
H.~H.~Duan, Y.~L.~Liu and M.~Q.~Huang,
"Light-cone sum rule analysis of semileptonic decays $\varLambda _b^0 \rightarrow \varLambda _c^+ \ell ^- {\overline{\nu }}_\ell $,''
\href{https://doi.org/10.1140/epjc/s10052-022-10931-8}{Eur. Phys. J. C \textbf{82}, no.10, 951 (2022)}.
\href{https://arxiv.org/pdf/2204.00409.pdf}{[arXiv:2204.00409 [hep-ph]]}.

\bibitem{Azizi:2018axf}
K.~Azizi and J.~Y.~S\"ung\"u,
"Semileptonic $\Lambda_{b}\rightarrow \Lambda_{c}{\ell}\bar\nu_{\ell}$ Transition in Full QCD,''
\href{https://doi.org/10.1103/PhysRevD.97.074007}{Phys. Rev. D \textbf{97}, no.7, 074007 (2018)}.
\href{https://arxiv.org/pdf/1803.02085.pdf}{[arXiv:1803.02085 [hep-ph]]}.



\bibitem{LHCb:2022piu}
R.~Aaij \textit{et al.} [LHCb],
"Observation of the decay $ \Lambda_b^0\rightarrow \Lambda_c^+\tau^-\overline{\nu}_{\tau}$,''
\href{https://doi.org/10.1103/PhysRevLett.128.191803}{Phys. Rev. Lett. \textbf{128}, no.19, 191803 (2022)}.
\href{https://arxiv.org/pdf/2201.03497.pdf}{[arXiv:2201.03497 [hep-ex]]}.

\bibitem{D0:2008sbw}
V.~M.~Abazov \textit{et al.} [D0],
"Observation of the doubly strange $b$ baryon $\Omega_b^-$,''
\href{https://doi.org/10.1103/PhysRevLett.101.232002}{Phys. Rev. Lett. \textbf{101}, 232002 (2008)}.
\href{https://arxiv.org/pdf/0808.4142.pdf}{[arXiv:0808.4142 [hep-ex]]}.

\bibitem{CDF:2009sbo}
T.~Aaltonen \textit{et al.} [CDF],
"Observation of the $\Omega_b^-$ and Measurement of the Properties of the $\Xi_b^-$ and $\Omega_b^-$,''
\href{https://doi.org/10.1103/PhysRevD.80.072003}{Phys. Rev. D \textbf{80}, 072003 (2009)}.
\href{https://arxiv.org/pdf/0905.3123.pdf}{[arXiv:0905.3123 [hep-ex]]}.

\bibitem{LHCb:2016coe}
R.~Aaij \textit{et al.} [LHCb],
"Measurement of the mass and lifetime of the $\Omega_b^-$ baryon,''
\href{https://doi.org/10.1103/PhysRevD.93.092007}{Phys. Rev. D \textbf{93}, no.9, 092007 (2016).}
\href{https://arxiv.org/pdf/1604.01412.pdf}{[arXiv:1604.01412 [hep-ex]]}.

\bibitem{Xu:1992hj}
Q.~P.~Xu and A.~N.~Kamal,
"Inclusive decays of bottom baryons,''
\href{https://doi.org/10.1103/PhysRevD.47.2849}{Phys. Rev. D \textbf{47}, 2849-2857 (1993).}

\bibitem{Cheng:1995fe}
H.~Y.~Cheng and B.~Tseng,
"1/M corrections to baryonic form-factors in the quark model,''
\href{https://doi.org/10.1103/PhysRevD.53.1457}{Phys. Rev. D \textbf{53}, 1457 (1996)}{[erratum: Phys. Rev. D \textbf{55}, 1697 (1997)]}.
\href{https://arxiv.org/pdf/hep-ph/9502391.pdf}{[arXiv:hep-ph/9502391 [hep-ph]]}.

\bibitem{Singleton:1990ye}
R.~L.~Singleton,
"Semileptonic baryon decays with a heavy quark,''
\href{https://doi.org/10.1103/PhysRevD.43.2939}{Phys. Rev. D \textbf{43}, 2939-2950 (1991)}.

\bibitem{Ivanov:1996fj}
M.~A.~Ivanov, V.~E.~Lyubovitskij, J.~G.~Korner and P.~Kroll,
"Heavy baryon transitions in a relativistic three quark model,''
\href{https://doi.org/10.1103/PhysRevD.56.348}{Phys. Rev. D \textbf{56}, 348-364 (1997)}.
\href{https://arxiv.org/pdf/hep-ph/9612463.pdf}{[arXiv:hep-ph/9612463 [hep-ph]]}.

\bibitem{Ebert:2006hm}
D.~Ebert, R.~N.~Faustov and V.~O.~Galkin,
"Relativistic description of semileptonic decays of heavy baryons,''
Conf. Proc. C \textbf{060726}, 1066-1069 (2006).
\href{https://arxiv.org/pdf/hep-ph/0610238.pdf}{[arXiv:hep-ph/0610238 [hep-ph]]}.

\bibitem{Ebert:2008oxa}
D.~Ebert, R.~N.~Faustov and V.~O.~Galkin,
"Properties of heavy baryons in the relativistic quark model,''
\href{https://inspirehep.net/files/bce60f6b7d7df643226ea8298076bd12}{[inspirehep.net/files/bce60f6b7d7df643226ea8298076bd12].}

\bibitem{Ebert:2006rp}
D.~Ebert, R.~N.~Faustov and V.~O.~Galkin,
"Semileptonic decays of heavy baryons in the relativistic quark model,''
\href{https://doi.org/10.1103/PhysRevD.73.094002}{Phys. Rev. D \textbf{73}, 094002 (2006)}.
\href{https://arxiv.org/pdf/hep-ph/0604017.pdf}{[arXiv:hep-ph/0604017 [hep-ph]]}.

\bibitem{Ivanov:1999pz}
M.~A.~Ivanov, J.~G.~Korner, V.~E.~Lyubovitskij, M.~A.~Pisarev and A.~G.~Rusetsky,
"On the choice of heavy baryon currents in the relativistic three quark model,''
\href{https://doi.org/10.1103/PhysRevD.61.114010}{Phys. Rev. D \textbf{61}, 114010 (2000).}
\href{https://arxiv.org/pdf/hep-ph/9911425.pdf}{[arXiv:hep-ph/9911425 [hep-ph]]}.

\bibitem{Pervin:2006ie}
M.~Pervin, W.~Roberts and S.~Capstick,
"Semileptonic decays of heavy omega baryons in a quark model,''
\href{https://doi.org/10.1103/PhysRevC.74.025205}{Phys. Rev. C \textbf{74}, 025205 (2006).}
\href{https://arxiv.org/pdf/nucl-th/0603061.pdf}{[arXiv:nucl-th/0603061 [nucl-th]]}.

\bibitem{Sheng:2020drc}
J.~H.~Sheng, J.~Zhu, X.~N.~Li, Q.~Y.~Hu and R.~M.~Wang,
"Probing new physics in semileptonic $\Sigma_b$ and $\Omega_b$ decays,''
\href{https://doi.org/10.1103/PhysRevD.102.055023}{Phys. Rev. D \textbf{102}, no.5, 055023 (2020)}.
\href{https://arxiv.org/pdf/2009.09594.pdf}{[arXiv:2009.09594 [hep-ph]]}.

\bibitem{Ivanov:1998ya}
M.~A.~Ivanov, J.~G.~Korner, V.~E.~Lyubovitskij and A.~G.~Rusetsky,
"Charm and bottom baryon decays in the Bethe-Salpeter approach: Heavy to heavy semileptonic transitions,''
\href{https://doi.org/10.1103/PhysRevD.59.074016}{Phys. Rev. D \textbf{59}, 074016 (1999).}
\href{https://arxiv.org/pdf/hep-ph/9809254.pdf}{[arXiv:hep-ph/9809254 [hep-ph]]}.

\bibitem{Rusetsky:1997id}
A.~G.~Rusetsky, M.~A.~Ivanov, J.~G.~Korner and V.~E.~Lyubovitskij,
"Weak decays of heavy baryons in the covariant quasipotential approach,''
\href{https://arxiv.org/pdf/hep-ph/9710524.pdf}{[arXiv:hep-ph/9710524 [hep-ph]]}.

\bibitem{Han:2020sag}
C.~Han and C.~Liu,
"b-baryon semi-tauonic decays in the Standard Model,''
\href{https://doi.org/10.1016/j.nuclphysb.2020.115262}{Nucl. Phys. B \textbf{961}, 115262 (2020).}
\href{https://arxiv.org/pdf/2011.00473.pdf}{[arXiv:2011.00473 [hep-ph]]}.

\bibitem{Du:2011nj}
M.~k.~Du and C.~Liu,
"$\Omega_b$ semi-leptonic weak decays,''
\href{https://doi.org/10.1103/PhysRevD.84.056007}{Phys. Rev. D \textbf{84}, 056007 (2011).}
\href{https://arxiv.org/pdf/1107.2535.pdf}{[arXiv:1107.2535 [hep-ph]]}.

\bibitem{Zhao:2018zcb}
Z.~X.~Zhao,
"Weak decays of heavy baryons in the light-front approach,''
\href{https://doi.org/10.1088/1674-1137/42/9/093101}{Chin. Phys. C \textbf{42}, no.9, 093101 (2018).}
\href{https://arxiv.org/pdf/1803.02292.pdf}{[arXiv:1803.02292 [hep-ph]]}.

\bibitem{Ke:2012wa}
H.~W.~Ke, X.~H.~Yuan, X.~Q.~Li, Z.~T.~Wei and Y.~X.~Zhang,
"$\Sigma_{b}\to\Sigma_c$ and $\Omega_b\to\Omega_c$ weak decays in the light-front quark model,''
\href{https://doi.org/10.1103/PhysRevD.86.114005}{Phys. Rev. D \textbf{86}, 114005 (2012).}
\href{https://arxiv.org/pdf/1207.3477.pdf}{[arXiv:1207.3477 [hep-ph]]}.

\bibitem{Xu:1993mj}
Q.~P.~Xu,
"A Bjorken sum rule for semileptonic Omega(b) decays to ground and excited charmed baryon states,''
\href{https:/doi.org/10.1103/PhysRevD.48.5429}{Phys. Rev. D \textbf{48}, 5429-5432 (1993).}
\href{https://arxiv.org/abs/hep-ph/9305349}{[arXiv:hep-ph/9305349 [hep-ph]]}.


\bibitem{Shifman:1978by}
M.~A.~Shifman, A.~I.~Vainshtein and V.~I.~Zakharov,
"QCD and Resonance Physics: Applications,''
\href{https://doi.org/10.1016/0550-3213(79)90023-3}{Nucl. Phys. B \textbf{147}, 448-518 (1979)}.

\bibitem{Shifman:1978bx}
M.~A.~Shifman, A.~I.~Vainshtein and V.~I.~Zakharov,
"QCD and Resonance Physics. Theoretical Foundations,''
\href{https://doi.org/10.1016/0550-3213(79)90022-1}{Nucl. Phys. B \textbf{147}, 385-447 (1979)}.


\bibitem{Shifman:2001ck}
M.~Shifman and B.~Ioffe,
"At the frontier of particle physics. Handbook of QCD. Vol. 1-3,''
World Scientific, 2001,
\href{https://doi.org/10.1142/4544}{ISBN 978-981-02-4445-3, 978-981-4492-22-5.}

\bibitem{Aliev:2010uy}
T.~M.~Aliev, K.~Azizi and M.~Savci,
"Analysis of the $\Lambda_{b}\rightarrow \Lambda \ell^+\ell^- $ decay in QCD,''
\href{https://doi.org/10.1103/PhysRevD.81.056006}{Phys. Rev. D \textbf{81}, 056006 (2010).}
\href{https://arxiv.org/pdf/1001.0227.pdf}{[arXiv:1001.0227 [hep-ph]]}.

\bibitem{Aliev:2009jt}
T.~M.~Aliev, K.~Azizi and A.~Ozpineci,
"Radiative Decays of the Heavy Flavored Baryons in Light Cone QCD Sum Rules,''
\href{https://doi.org/10.1103/PhysRevD.79.056005}{Phys. Rev. D \textbf{79}, 056005 (2009).}
\href{https://arxiv.org/pdf/0901.0076.pdf}{[arXiv:0901.0076 [hep-ph]]}.

\bibitem{Aliev:2012ru}
T.~M.~Aliev, K.~Azizi and M.~Savci,
"Doubly Heavy Spin--1/2 Baryon Spectrum in QCD,''
\href{https://doi.org/10.1016/j.nuclphysa.2012.09.009}{Nucl. Phys. A \textbf{895}, 59-70 (2012).}
\href{https://arxiv.org/pdf/1205.2873.pdf}{[arXiv:1205.2873 [hep-ph]]}.

\bibitem{Agaev:2016dev}
S.~S.~Agaev, K.~Azizi and H.~Sundu,
"Strong $Z_c^{+}(3900)\rightarrow J/\psi \pi^{+}; \eta_{c} \rho^{+}$ decays in QCD,''
\href{https://doi.org/10.1103/PhysRevD.93.074002}{Phys. Rev. D \textbf{93}, no.7, 074002 (2016).}
\href{https://arxiv.org/pdf/1601.03847.pdf}{[arXiv:1601.03847 [hep-ph]]}.

\bibitem{Azizi:2016dhy}
K.~Azizi, Y.~Sarac and H.~Sundu,
"Analysis of $P_c^+(4380)$ and $P_c^+(4450)$ as pentaquark states in the molecular picture with QCD sum rules,''
\href{https://doi.org/10.1103/PhysRevD.95.094016}{Phys. Rev. D \textbf{95}, no.9, 094016 (2017).}
\href{https://arxiv.org/pdf/1612.07479.pdf}{[arXiv:1612.07479 [hep-ph]]}.



\bibitem{Shifman:2010zzb}
M.~Shifman,
"Vacuum structure and QCD sum rules: Introduction,''
\href{https://doi.org/10.1142/S0217751X1004855X}{Int. J. Mod. Phys. A \textbf{25}, 226-235 (2010)}.

\bibitem{Gross:2022hyw}
F.~Gross, E.~Klempt, S.~J.~Brodsky, A.~J.~Buras, V.~D.~Burkert, G.~Heinrich, K.~Jakobs, C.~A.~Meyer, K.~Orginos and M.~Strickland, \textit{et al.}
"50 Years of Quantum Chromodynamics,''
\href{https://doi.org/10.1140/epjc/s10052-023-11949-2}{Eur. Phys. J. C \textbf{83}, 1125 (2023).}
\href{https://arxiv.org/ftp/arxiv/papers/2212/2212.11107.pdf}{[arXiv:2212.11107 [hep-ph]]}.


\bibitem{Agaev:2017jyt}
S.~S.~Agaev, K.~Azizi and H.~Sundu,
"On the nature of the newly discovered $\Omega$ states,''
\href{https://doi.org/10.1209/0295-5075/118/61001}{EPL \textbf{118}, no.6, 61001 (2017).}
\href{https://arxiv.org/pdf/1703.07091.pdf}{[arXiv:1703.07091 [hep-ph]]}.

\bibitem{Agaev:2020zad}
S.~Agaev, K.~Azizi and H.~Sundu,
"Four-quark exotic mesons,''
\href{https://doi.org/10.3906/fiz-2003-15}{Turk. J. Phys. \textbf{44}, no.2, 95-173 (2020).}
\href{https://arxiv.org/pdf/2004.12079.pdf}{[arXiv:2004.12079 [hep-ph]]}.


\bibitem{Azizi:2017ubq}
K.~Azizi and N.~Er,
"X (3872): propagating in a dense medium,''
\href{https://doi:10.1016/j.nuclphysb.2018.09.014}{Nucl. Phys. B \textbf{936}, 151-168 (2018)}.
\href{https://arxiv.org/pdf/1710.02806.pdf}{[arXiv:1710.02806 [hep-ph]]}.

\bibitem{Aliev:2006gk}
T.~M.~Aliev, K.~Azizi and A.~Ozpineci,
"Semileptonic $B(s) \to D(sJ)(2460)l \nu$ decay in QCD,''
\href{https://doi.org/doi:10.1140/epjc/s10052-007-0315-6}{Eur. Phys. J. C \textbf{51}, 593-599 (2007)}.
\href{https://arxiv.org/pdf/hep-ph/0608264.pdf}{[arXiv:hep-ph/0608264 [hep-ph]]}.




\bibitem{ParticleDataGroup:2020ssz}
P.~A.~Zyla \textit{et al.} [Particle Data Group],
"Review of Particle Physics,''
\href{https://academic.oup.com/ptep/article/2020/8/083C01/5891211?login=true}{PTEP \textbf{2020}, no.8, 083C01 (2020)}.

\bibitem{Belyaev:1982sa}
V.~M.~Belyaev and B.~L.~Ioffe,
"Determination of Baryon and Baryonic Resonance Masses from QCD Sum Rules. 1. Nonstrange Baryons,''
\href{http://www.jetp.ac.ru/cgi-bin/dn/e_056_03_0493.pdf}{Sov. Phys. JETP \textbf{56}, 493-501 (1982)}.
ITEP-59-1982.

\bibitem{Belyaev:1982cd}
V.~M.~Belyaev and B.~L.~Ioffe,
"Determination of the baryon mass and baryon resonances from the quantum-chromodynamics sum rule. Strange baryons,''
\href{http://www.jetp.ac.ru/cgi-bin/dn/e_057_04_0716.pdf}{Sov. Phys. JETP \textbf{57}, 716-721 (1983)}.
ITEP-132-1982.

\bibitem{Ioffe:2005ym}
B.~L.~Ioffe,
"QCD at low energies,''
\href{https://doi.org/10.1016/j.ppnp.2005.05.001}{Prog. Part. Nucl. Phys. \textbf{56}, 232-277 (2006)}.
\href{https://arxiv.org/pdf/hep-ph/0502148.pdf}{[arXiv:hep-ph/0502148 [hep-ph]]}.



\bibitem{Migura:2006en}
S.~Migura, D.~Merten, B.~Metsch and H.~R.~Petry,
"Semileptonic decays of baryons in a relativistic quark model,''
\href{https://doi.org/10.1140/epja/i2006-10024-x}{Eur. Phys. J. A \textbf{28}, 55 (2006).}
\href{https://arxiv.org/pdf/hep-ph/0602152.pdf}{[arXiv:hep-ph/0602152 [hep-ph]]}.

\bibitem{Korner:1994nh}
J.~G.~Korner, M.~Kramer and D.~Pirjol,
"Heavy baryons,''
Prog. Part. Nucl. Phys. \textbf{33}, 787-868 (1994)
\href{https://doi.org/10.1016/0146-6410(94)90053-1}{Prog. Part. Nucl. Phys. \textbf{33}, 787-868 (1994).}
\href{https://arxiv.org/pdf/hep-ph/9406359.pdf}{[arXiv:hep-ph/9406359 [hep-ph]]}.


\bibitem{Bialas:1992ny}
P.~Bialas, J.~G.~Korner, M.~Kramer and K.~Zalewski,
"Joint angular decay distributions in exclusive weak decays of heavy mesons and baryons,''
\href{https://doi.org/10.1007/BF01555745}{Z. Phys. C \textbf{57}, 115-134 (1993)}.
%
%



  
  






  

\end{thebibliography}
\end{document}